\documentclass[traditabstract]{aa} 
\usepackage{graphicx}
\usepackage{txfonts}
\usepackage{color}
\graphicspath{{./figures_final/}}
\DeclareRobustCommand{\ion}[2]{\textup{#1\,\textsc{\lowercase{#2}}}}
\DeclareRobustCommand{\us}{\,}

\begin{document} 
%%%%%%%%%%%%%%%%%%%%%%%%%%%%%%%%%%%%%%%%%
% Title & Affiliations & Abstract
%%%%%%%%%%%%%%%%%%%%%%%%%%%%%%%%%%%%%%%%%

% Title & Affiliations
  \title{X-Shooter observations of low-mass stars in the\\$\eta$~Chamaeleontis association\thanks{This work is based on observations made with ESO Telescopes at the Paranal Observatory under program ID 084.C-1095. }}

%  \subtitle{XXX}
  \author{Michael Rugel
     \inst{1}
     \and
     Davide Fedele
     \inst{2}
     \and
     Gregory Herczeg
     \inst{3}     
     }

  \institute{Max Planck Institute for Astronomy, K\"onigstuhl 17, 69117 Heidelberg, Germany\\
       \email{rugel@mpia.de}
     \and
       INAF--Osservatorio Astrofisico di Arcetri, L.go E. Fermi 5, 50125 Firenze, Italy
     \and
       Kavli Institute for Astronomy and Astrophysics, Peking University, Yi He Yuan Lu 5, Haidian Qu, 100871 Beijing, China
       }
 \abstract
 {The nearby $\eta$~Chamaeleontis association is a collection of 4--10{\us}Myr old stars with a disk 
 fraction of 35--45\%. In this study, the broad wavelength coverage of VLT/X-Shooter is used to
 measure the stellar and mass accretion properties of 15 low mass
 stars in the $\eta$~Chamaeleontis association. For each star, the observed spectrum
 is fitted with a non-accreting stellar template and an accretion spectrum 
 obtained from assuming a plane-parallel hydrogen slab. 
 Five of the eight stars with an IR disk excess show excess UV emission, indicating ongoing accretion.
 The accretion rates measured here are similar to those obtained from previous measurements 
 of excess UV emission, but tend to be higher than past measurements from H$\alpha$ modeling.
 The mass accretion rates are consistent with
 those of other young star forming regions. 
}
  \date{Received XXX; accepted XXX}
  \maketitle

%%%%%%%%%%%%%%%%%%%%%%%%%%%%%%%%%%%%%%%
% Introduction
%%%%%%%%%%%%%%%%%%%%%%%%%%%%%%%%%%%%%%%
\section{Introduction}	   \label{sec:introduction}
The evolution of protoplanetary disks affects planet formation and migration. 
Statistical studies of both excess infrared and H$\alpha$ emission demonstrate that the majority of disks
dissipate within 2--3{\us}Myrs after the collapse of the parent
molecular cloud
\cite[e.g.,][]{HaischLada:2001aa,Fedelevan-den-Ancker:2010aa}. This
timescale sets a limit on the time available to build up the
atmospheres of gas giant planets. The evolution and dissipation of
disks differs from system to system -- a fraction
of stars retain their gas and dust disks for 5--10{\us}Myrs.

\smallskip
\noindent
One such cluster, the $\eta$~Chamaeleontis association \citep{MamajekLawson:1999aa}, is
located at a distance of $d=94${\us}pc\footnote{Recent parallax measurements with
\textsc{gaia} give distances of 94 and 102{\us}pc for HD 75505 and RECX~1, 
respectively \citep{Gaia-CollaborationBrown:2016aa}, in good agreement with previous estimates. 
In this paper we adopt a distance of 94{\us}pc.} (\citealt{van-Leeuwen:2007aa}; see discussion in 
\citealt{MurphyLawson:2010aa}) with an age of \mbox{4$-$10{\us}Myr}
\citep[e.g.,][]{LawsonCrause:2001aa,LawsonFeigelson:2001aa,LuhmanSteeghs:2004aa,HerczegHillenbrand:2015aa}. The total disk 
fraction of the association ranges between 35$-$45\% if including stars of all masses\footnote{The lower bound takes into account also 
7 recent $\eta$~Cha cluster member candidates
\citep[][]{MurphyLawson:2010aa,Lopez-MartiJimenez-Esteban:2013aa}, of which one has been confirmed to have a disk \citep{SimonSchlieder:2012aa}.}.
Of the 15 canonical low-mass members, 8 retain dust disks as identified from excess IR emission
\citep{MegeathHartmann:2005aa,Sicilia-AguilarBouwman:2009aa}, 
thereby offering a unique opportunity to study disk and stellar 
properties at stages when giant planet formation should be coming to an end. 

\smallskip
\noindent
In the viscous accretion model of disk evolution
\citep[e.g.][]{HartmannCalvet:1998aa}, the accretion rate is expected to
decrease with age. The accretion properties of old disks, including
those in the $\eta$~Chamaeleontis association, are therefore important
to compare to the accretion properties of younger systems.  However,
these measurements may be complicated because emission from young stars include photospheric and accretion
components and are affected by extinction, all of which may vary with time
\citep[e.g.,][]{BertoutBasri:1988aa,BasriBertout:1989aa,GullbringHartmann:1998aa,Sicilia-AguilarHenning:2010aa}.
Recent improvements in evaluating photospheric and accretion
properties of young stars have been driven by simultaneous fits of
spectral type, extinction, and accretion luminosity to broadband
optical spectra, using young stars as templates \citep[e.g.,][]{ManaraBeccari:2013aa,ManaraFedele:2016aa,ManaraTesti:2017aa,HerczegHillenbrand:2014aa}.

\smallskip
\noindent
In this paper, we analyze flux-calibrated X-Shooter optical spectra of
15 low-mass members of the $\eta$ Cha cluster to measure accretion and
photospheric properties of stars in the association.
The paper is structured as follows: After describing the observational setup in Section~2, in
Section~3 we determine the stellar parameters and investigate each
star for extinction and continuum excess; in Section~4, we investigate each
sample member for accretion and infer the accretion luminosity of the
accreting PMSs; in Section~5 we present hydrogen emission
lines and the derived mass accretion rates. In Section~6 we discuss
the results and summarize the study in Section~7.  

%%%%%%%%%%%%%%%%%%%%%%%%%%%%%%%%%%%%%%
% Data reduction
%%%%%%%%%%%%%%%%%%%%%%%%%%%%%%%%%%%%%%

\begin{table*}
\caption{Observation log of VLT/X-Shooter observations in the $\eta$~Cha association} 
\centering
\begin{tabular}{lrrrlrrr}
 \hline
 \hline
 Object &RA(J2000)&DEC(J2000)&Obs. date&Mode&Exp. (UVB)&Exp. (VIS)&Exp. (NIR)\\
 &&&[UTC]&&[s]&[s]&[s]\\
 \hline
 J0836.2-7908 (J0836) &   8 36 10.27&  $-$79 08 17.65&	2010-01-20&	NOD &800	&790 &780 \\
  RECX~1		&   8 36 55.77&  $-$78 56 45.71&	2010-01-18&	NOD & 48	& 20 & 40 \\
 J0838.9-7916 (J0838) &   8 38 50.65&  $-$79 16 13.66&	2010-01-18&	NOD &200	&180 &200 \\
 J0841.5-7853 (J0841) &   8 41 29.24&  $-$78 53 03.45&	2010-01-19&	NOD &360	&388 &388 \\
  RECX~3		&   8 41 36.46&  $-$79 03 27.97&	2010-01-20&	NOD &220	&240 &240 \\
  RECX~4		&   8 42 23.07&  $-$79 04 00.60&	2010-01-18&	NOD &100	& 90 & 80 \\
  RECX~5        &   8 42 26.37&  $-$78 57 44.48&	2010-01-19&	NOD &120	& 96 &120 \\
  RECX~6		&   8 42 38.71&  $-$78 54 41.97&	2010-01-20&	NOD &170	&180 &180 \\
  RECX~7		&   8 43 07.19&  $-$79 04 50.84&	2010-01-20&	STARE& 55	& 60 & 60 \\
 J0843.3-7915 (J0843) &   8 43 17.24&  $-$79 05 16.74&	2010-01-18&	NOD &100	& 90 &100 \\
 J0844.2-7833 (J0844) &   8 44 08.61&  $-$78 33 45.25&	2010-01-18&	NOD &460	&480 &460 \\
 RECX~9		&   8 44 15.65&  $-$78 59 05.43&	2010-01-19&	NOD &240	& 60 &240 \\
 RECX~10		&   8 44 30.81&  $-$78 46 29.02&	2010-01-19&	NOD &100	& 90 &100 \\
 RECX~11		&   8 47 01.25&  $-$78 59 34.02&	2010-01-18&	NOD & 50	& 30 & 40 \\
 RECX~12		&   8 47 55.72&  $-$78 54 52.74&	2010-01-19&	NOD &100	&120 &120 \\
 \hline
\end{tabular}
\tablefoot{
The first column lists the names of all objects in the observed sample. Abbreviations, which have been used in this work, are given in parenthesis. 
 The observations 
 were carried out in ``NODDING'' (NOD)/``STARE'' mode with the $1\arcsec\times11\arcsec$ slit in
 the UVB arm and $0\farcs4\times11\arcsec$ slits in both VIS and NIR arm. The exposure time (Exp.) for each arm is given in the last three columns. Complementary
 broad slit observations in ``STARE'' mode used the $5\arcsec\times11\arcsec$ slits for UVB and VIS arm, as
 well as $1\farcs5\times11\arcsec$ slit for the NIR arm (except for targets J0836, RECX~3,
 RECX~6 and RECX~7, for which a slit of $5\arcsec\times11\arcsec$ was used in the NIR
 arm).
}
\label{tbl:obslog_etacha}
\end{table*}

\section{Observations and data reduction} \label{sec:observations}
As part of a survey on T Tauri stars in the $\eta$~Cha association and Chamaeleon I and II (\citealt{ManaraFedele:2016aa}; Program ID 084.C-1095, PI: Herczeg), a sub-sample of $\eta$~Cha cluster members 
have been observed with the ESO/VLT X-Shooter echelle spectrograph over three nights in January 2010 (Table~\ref{tbl:obslog_etacha}).
The X-Shooter sample consists of 15 low-mass stars among the cluster members found by, e.g., \citet{LuhmanSteeghs:2004aa}. Four new probable cluster members and three potential members were identified by \citet{MurphyLawson:2010aa} in the outskirts of the cluster, after the observations presented here had been conducted, and are therefore not included in our sample. 

\smallskip
\noindent
The X-Shooter spectrograph covers a broad wavelength range, from 300 nm to 2500 nm using three different arms (UVB: $\lambda\lambda$~300--550{\us}${\rm nm}$, VIS:
$\lambda\lambda$~550--1000{\us}${\rm nm}$, NIR: $\lambda\lambda$~1000--2500{\us}${\rm nm}$, \citep{VernetDekker:2011aa}). 
To optimize the flux calibration, we combined short (from 3--65 seconds) broad-slit ($1\farcs5 - 5\farcs0$) and deep (from 20--800 seconds) narrow-slit ($0\farcs4-1\farcs0$) observations. The observation log
is given in Table \ref{tbl:obslog_etacha}. In this paper we present the optical spectra extracted from the UVB and VIS arms. 

\smallskip
\noindent
Data reduction, carried out with the X-Shooter pipeline XSH~1.2.0
\citep{ModiglianiGoldoni:2010aa}, consisted of bias and flat-field
corrections, combination of single frames obtained in the ``NODDING''
mode, wavelength calibration, rectifying each order, and then merging the
orders. The extraction of the spectra and the sky removal were performed with the IRAF~\footnote{IRAF is distributed by the National Optical Astronomy Observatories,
 which are operated by the association of Universities for Research in Astronomy, Inc, under cooperative agreement with the National Science Foundation.} task \texttt{apall}. 

\smallskip
\noindent
The response function was measured using the \mbox{spectro-photometric} standard star GD~71 \citep{HamuySuntzeff:1994aa,VernetKerber:2009aa} observed at the beginning of each night. 
Since the comparison of the response functions from the individual nights only yielded small differences, all UVB-arm observations were calibrated with the response function
derived from the first night and all VIS-arm observations with an average response function of all three nights.

\begin{figure*}	%[ht]	
\centering			
\includegraphics[width=.85\linewidth]{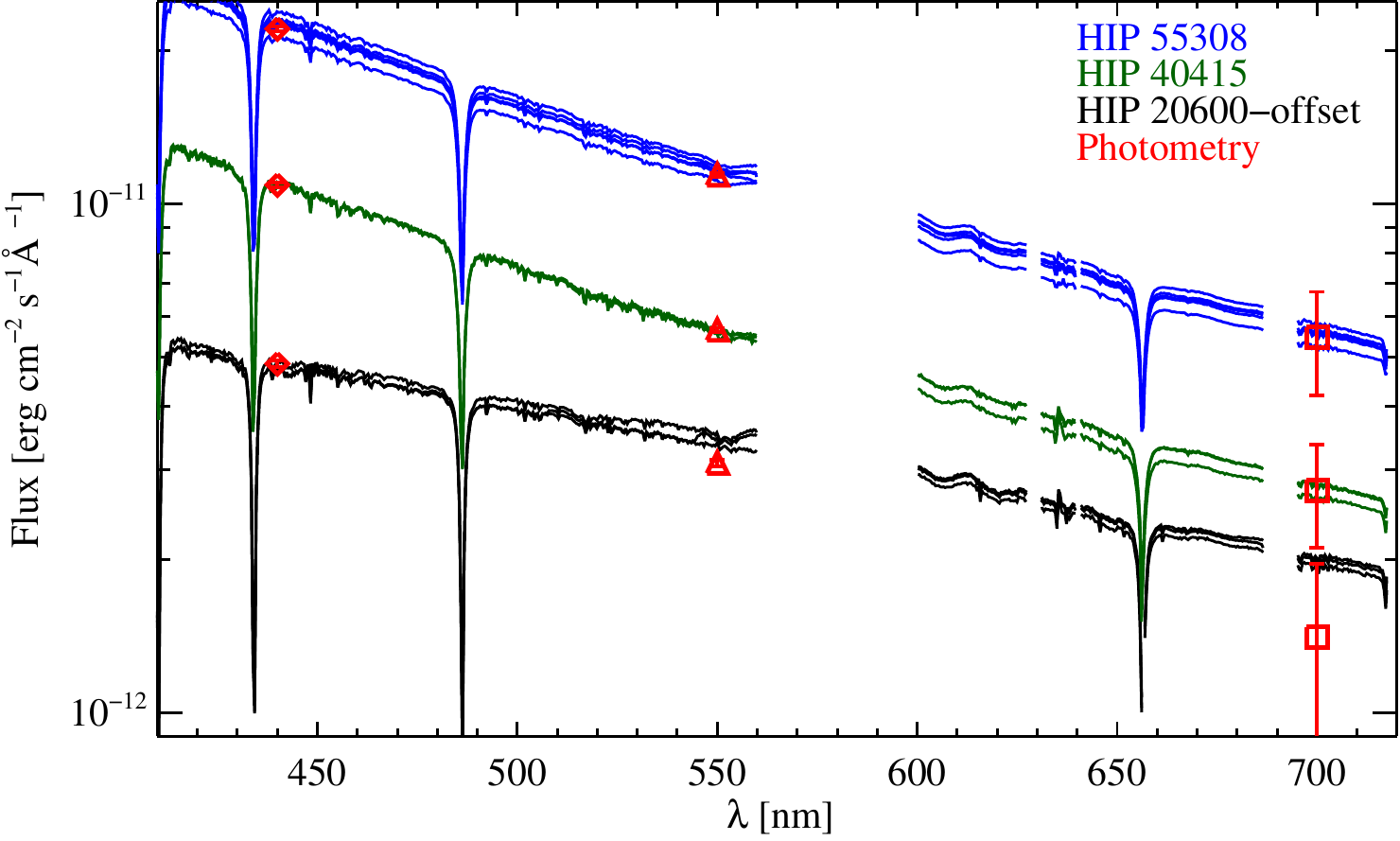}
\caption{Comparison of telluric standards HIP 55308 (blue), HIP 40415 (green) and HIP 20600 (black), observed in different epochs (multiple instances of one color), with literature photometry (red) in the $B$-band (diamonds), $V$-band (triangles) and $R$-band (squares). The spectra of HIP 20600 and the corresponding photometry are displayed with a constant offset of $1\times10^{-12}\,{\rm erg\,s^{-1}\,cm^{-2}\,\AA^{-1}}$.}
\label{fig:plot_app_plot_phot}
\end{figure*}

\smallskip
\noindent
In order to account for wavelength-dependent slit losses, \mbox{narrow-slit} observations are scaled 
to the \mbox{broad-slit} observations with low order polynomials fitted to spectral regions 
with high signal to noise.
The stability and accuracy of the flux calibration is estimated by comparing observations of telluric standards across all nights. 
Figure \ref{fig:plot_app_plot_phot} shows the comparison of \mbox{multi-epoch} X-Shooter spectra of three standards stars with
optical photometry: $B_T,V_T$ band photometry from Tycho-2 catalogue\footnote{http://cdsarc.u-strasbg.fr/viz-bin/Cat?I/259; converted to Johnson $B$ and $V$-band photometry} 
and $R$~band from the NOMAD catalogue\footnote{http://cdsarc.u-strasbg.fr/viz-bin/Cat?I/297; $R$-band photometry for HIP 40415 from USNO-B1.0 catalog, http://cdsarc.u-strasbg.fr/viz-bin/Cat?I/284}.  In
general, the observed spectra agree well with the photometry, with
fluxes at $B$, $V$, and $R$-band wavelengths located within the error bars
of the photometry and an overall scatter in all standards of $\sim$$4\%$. Based on this comparison the flux calibration accuracy is $\sim$$4\%$.

%%%%%%%%%%%%%%%%%%%%%%%%%%%%%%%%%%%%%%
% Stellar Parameters
%%%%%%%%%%%%%%%%%%%%%%%%%%%%%%%%%%%%%%
\section{Stellar parameters} \label{sec:stellarparameters}
Measurements of stellar parameters for low-mass pre-main sequence
stars may be contaminated by the accretion continuum, which if not considered can lead to 
degeneracies between spectral type, extinction, and 
accretion measurements. 
These degeneracies are minimized when measuring stellar parameters from flux-calibrated optical spectra that 
cover a wide wavelength range \citep[e.g.,][]{ManaraBeccari:2013aa,HerczegHillenbrand:2014aa,AlcalaNatta:2014aa}. 
In this section we first estimate spectral type (SpT) and veiling
following \citet{HerczegHillenbrand:2014aa}. Based on this initial
estimate, we then perform a grid comparison of the program stars with stellar templates of different SpT, and different veiling and extinction values. 
Finally, the physical stellar properties (effective temperature ($T_{\rm eff}$), luminosity ($L_{*}$), radius ($R_{*}$) and mass ($M_{*}$))
are inferred using
tabulated conversions, and from
comparison with evolutionary tracks on the Hertzsprung-Russell diagram. 

\subsection{Spectral types} \label{sec:stpa_spt}
Our initial spectral types are calculated from atomic absorption and molecular
bands, quantified in the spectral indices
R5150, TiO6800, TiO7140 and TiO8465 that were developed for young
stars \citep[][]{HerczegHillenbrand:2014aa}. The indices TiO6800, TiO7140 and TiO8465 compare 
integrated flux within the absorption and the continuum band. For R5150, a 
linear fit over low and high wavelength continuum regions is used to
estimate the continuum on the band.  In two cases, J0843 and J0844,
these indices are corrected for veiling (see Sect.~3.2).  These spectral
types agree with those that would have been measured from the TiO7140
index of \citet{JeffriesOliveira:2007aa}. 

\smallskip
\noindent
Table~\ref{tbl:spt_ind_star} lists the spectral type obtained from
each index, which combine to form an initial estimate of spectral type. The final spectral types are adopted in Sect.~\ref{sec:stellarparametersgridcomparison} by comparing the spectra to a grid of spectra of non-accreting young stars. 

\begin{table*}%[ht]
\caption{Spectral types derived from spectral indices and adopted values after comparison to non-accreting PMS}
\centering	
\begin{tabular}{lccccccc} 
\hline \hline
Object  &	R5150   &  TiO6800  & TiO7140 & TiO8465  & JTiO7140   &Adopted\\	
	   &	(K0-M0)  &  (K5-M0.5) & (M0-M4.5)& (M4-M8)  & (K5-M7)   &    \\	
\hline
    J0836 & -     & -      & -     & M5.70  & M6.4     & M5.5\\
    RECX~1& K5.7   & K5.4    & -     & -    & K6.4     & K6.0\\
    J0838 & -     & -      & -     & M5.3   & M5.9     & M5.5\\
    J0841 & -     & -      & -     & M5.1   & M5.6     & M5.0\\
    RECX~3 & -     & -      & M3.60   & -    & M3.7     & M3.5\\
    RECX~4& -     & -    & M1.7   & -    & M1.3     & M1.5\\
    RECX~5& -     & -      & M4.5   &  M4.2  & M4.5     & M4.5\\
    RECX~6& -     & -      & M3.1   & -    & M3.2     & M3.0\\
    RECX~7& K5.6   & K6.0    & -     & -    & K6.9     & K6.0\\
    J0843 & -     & -      & M3.6   & -  & M3.7     & M4.0\\
    J0844 & -     & -      & -     & M6.1   & M5.5     & M6.0\\
    RECX~9& -     & -      & -     & M4.7   & M5.2     & M4.5\\
    RECX~10& -     & K8.4    & M0.7   & -    & M0.2     & M0.0\\
    RECX~11& K5.7   & K5.2    & -     & -    & K6.0     & K6.0\\
    RECX~12& -     & -      & M3.0   & -    & M3.1     & M3.0\\
\hline
\end{tabular}
\tablefoot{Columns 2--5 give spectral types from indices by \citet{HerczegHillenbrand:2014aa}. The index JTiO 7140 is listed as comparison \citep{JeffriesOliveira:2007aa}, however is not used to in the final classification. The validity range of each spectral index is indicated in brackets. The last column shows the adopted spectral type in this work (see Sects.~\ref{sec:stpa_spt} and \ref{sec:stellarparametersgridcomparison} for details).}
\label{tbl:spt_ind_star}
\end{table*}

\subsection{Veiling estimates} \label{sec:veiling}
\begin{figure}%[ht]
\centering
\includegraphics[scale=.85]{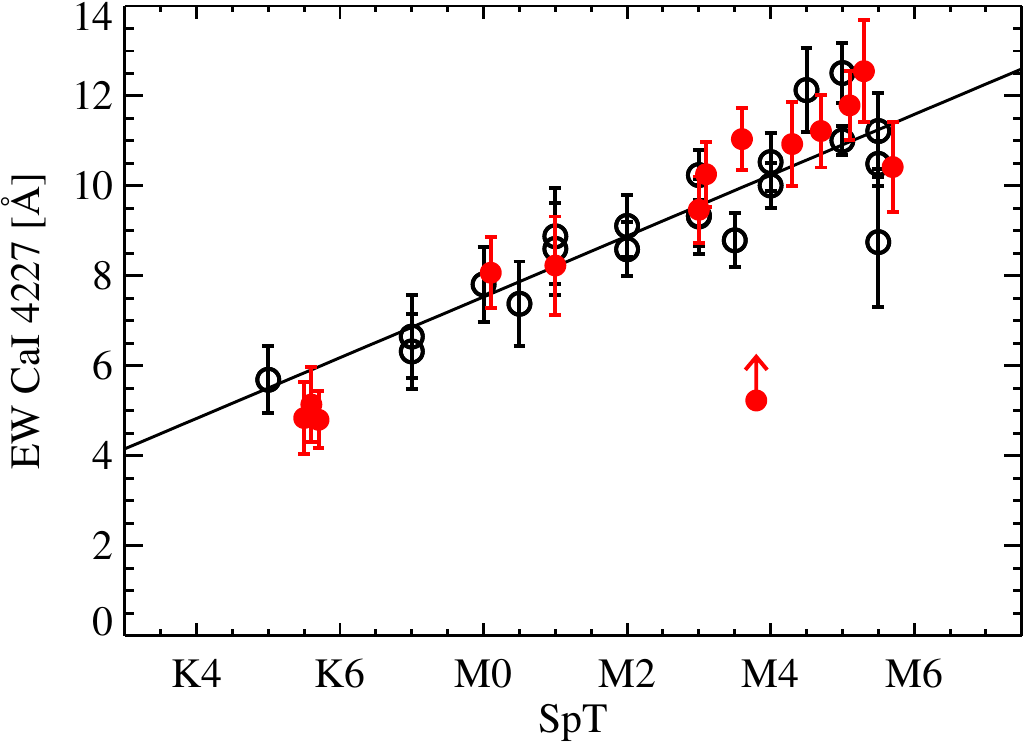}
\caption{
  Equivalent width of \ion{Ca}{i} $\lambda 422.7${\us}nm line vs. spectral type. Red filled
  circles denote observations presented in this work, black empty circles are from template stars
  \citep{ManaraTesti:2013aa}. Due to strong emission lines inside the absorption feature, 
  we determine a lower limit for J0843 (red arrow).
  The black line denotes the relation between the
  equivalent width and spectral type of a non-accreting stars (Eq.~\ref{eqn:ew_caI}).}
\label{fig:plot_veiling_ewcaI_vs_spt}
\end{figure}

Veiling describes the change of the depth of photospheric absorption
lines due to excess continuum emission
\citep[e.g][]{BasriBatalha:1990aa}. 
Veiling at blue wavelengths is measured in the gravity-sensitive
\ion{Ca}{i}~$\lambda 422.7$~nm absorption line, following the approach
of \citet{HerczegHillenbrand:2014aa} but adapted to higher spectral
resolution using the non-accreting photospheric templates from
\citet{ManaraTesti:2013aa}.  For templates, the relationship between spectral type and
equivalent width (Fig.~\ref{fig:plot_veiling_ewcaI_vs_spt}) is
described by
\begin{equation}
 \mathrm{EW}_{\mathrm{CaI}~\lambda \mathrm{422.7 nm}}=-31.7+0.676\times \mathrm{SpT},
\label{eqn:ew_caI}
\end{equation}
\noindent
where SpT is spectral type and a spectral type of M0 is equivalent to ${\rm SpT}=58$. 
Fig.~\ref{fig:plot_veiling_ewcaI_vs_spt} shows that most $\eta$-Cha-cluster 
objects have equivalent widths consistent with those expected from the
templates. These objects are assumed to have negligible veiling. 

\begin{figure}%[ht]
\centering
\includegraphics[scale=.78]{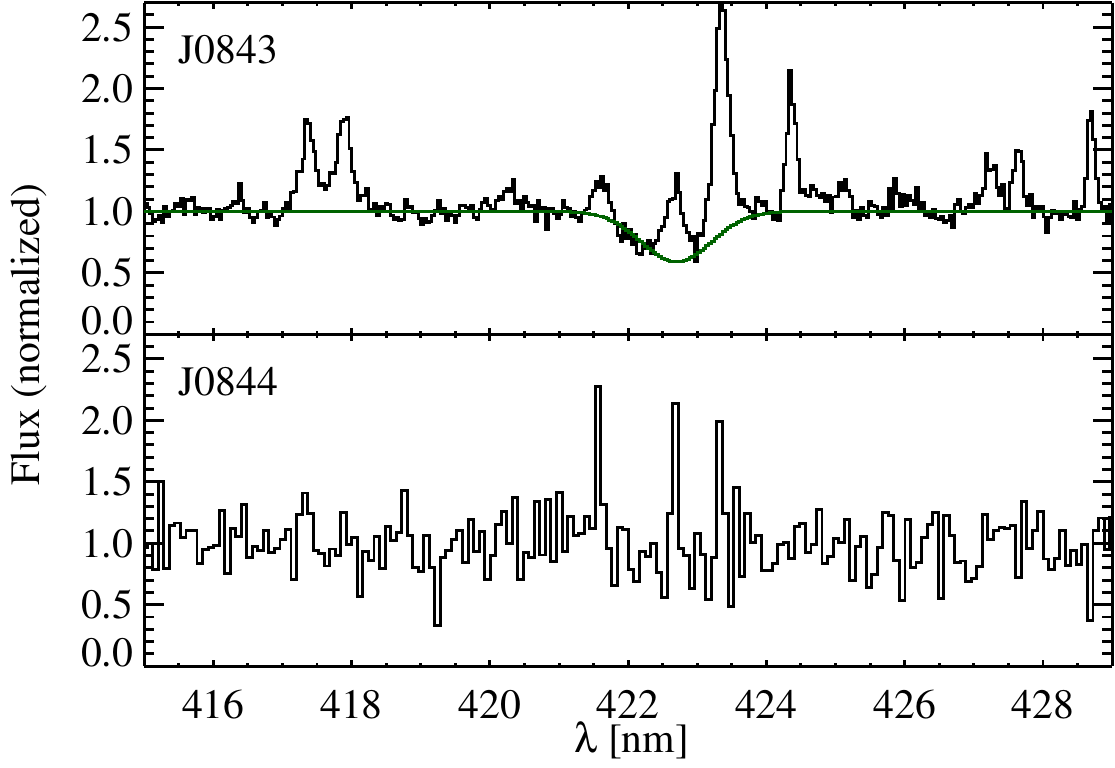}
\caption{
  Continuum-normalized spectrum around the  \ion{Ca}{i}~$\lambda
  422.7$~nm absorption line (black) for the veiled stars J0843 (top) and J0844
  (bottom).  The absorption line from J0843 is contaminated by
  emission in the same transition. The equivalent width, and subsequently the veiling, are
  measured by fitting a Gaussian profile (green) to the spectrum,
  avoiding regions with emission.  The absorption line from J0844 is
  not detected.
}
\label{fig:plot_veiling_J0843_gauss}
\end{figure}

\smallskip
\noindent
Two objects, J0843 and J0844, show significant veiling. The \mbox{\ion{Ca}{i} $\lambda 422.7${\us}nm} 
absorption feature in J0843 is detected but blended with emission lines. 
Assuming that the absorption profile is Gaussian with similar width as the template spectra, 
the lower limit on the \ion{Ca}{i} equivalent width is estimated to be at 5.2{\us}\AA\ (Fig.~\ref{fig:plot_veiling_J0843_gauss}), 
implying a veiling\footnote{The amount of veiling at wavelength $\lambda$ is given by \mbox{$r_\lambda = f_{\rm excess}/f_{\rm phot}$}, where $f_{\rm excess}(\lambda)$ is flux in excess to the flux of the photosphere, $f_{\rm phot}(\lambda)$.} of $r_{\mathrm 422.7{\us}{\rm nm}} \lesssim 1$ (Eq.~\ref{eqn:ew_caI}). The
absorption feature is not detected in J0844 
(Fig.~\ref{fig:plot_veiling_J0843_gauss}), indicating that the veiling
is high. 

\subsection{Grid comparison with PMS stellar templates} \label{sec:stellarparametersgridcomparison}
In order to verify the spectral type and veiling, we compare each
spectrum to a set of pre-main sequence, non-accreting stellar
templates obtained from the X-Shooter library of young stars \citep{ManaraTesti:2013aa}.  
We model veiling and extinction simultaneously
by adding constant accretion continuum flux to the template
\citep{HerczegHillenbrand:2014aa}, and by convolving each model with
the extinction curve of \citet{CardelliClayton:1989aa}, with a
total-to-selective extinction of $R_V=3.1$.  The best-fit parameters
are initially determined from spectral fits and are then tweaked
manually to obtain the best visual representation of the spectrum.

\smallskip
\noindent
All objects without veiling in the \ion{Ca}{i}~$\lambda 422.7${\us}nm
absorption feature (see Sect.~\ref{sec:veiling}) are best modeled
without adding excess flux longward of 400{\us}nm.  All sources,
including those with ongoing accretion, are
also well fit with no reddening, consistent with previous
studies \citep[e.g.,][]{LuhmanSteeghs:2004aa}. 

\smallskip
\noindent
An iterative approach is used for the two objects with strong veiling
in the \ion{Ca}{i} $\lambda 422.7$ line, J0843 and J0844. For the
final models, the extinction is fixed at $A_V = 0${\us}mag. The
templates are fixed to the pre-main sequence stars TWA 9B (TW Hya
association, SpT = M3) and SO925 ($\sigma$ Ori region, SpT = M5.5;
\citealt{ManaraTesti:2013aa}). The best-fit value for the veiling at
751 nm is $r_{751}=0.05$ in both objects. 
The veiling
translates to excess fluxes of $f_{\rm excess} = 9.1\times
10^{-16}\,{\rm erg\,s^{-1}\,cm^{-2}\,\AA^{-1}}$ for J0843 and $f_{\rm
  excess} = 6.9\times 10^{-17}\,{\rm erg\,s^{-1}\,cm^{-2}\,\AA^{-1}}$
for J0844. After accounting for this veiling, the spectral type from
spectral indices changes to M4.0 for  J0843 and stays at M6.0 for J0844.  

\smallskip
\noindent
The final spectral types are listed to a precision of one subclass for stars earlier than M0 and 
to 0.5 subclasses for stars later than M0, adopting the spacing of the sample of non-accreting template stars \citep[see also][]{ManaraTesti:2013aa}.

\subsection{Effective temperature, luminosity and stellar radius}  \label{sec:stpa_effectivetemperatureandluminosity}

\begin{table*}
\centering	
\caption{
 Stellar parameters - Literature comparison
}
\begin{tabular}{l|r|r|r|r|r|r|r|r} 
 \hline 
 \hline 
 Object & SpT &SpT &SpT &$T_{\rm eff}$ [K] &$T_{\rm eff}$ [K] &$L_{\ast}$ [$L_{\odot}$]& $L_{\ast}$ [$L_{\odot}$]&$L_{\ast}$ [$L_{\odot}$] \\
 & (Luhman\tablefootmark{1}) &(Lyo\tablefootmark{2}) & & (Luhman\tablefootmark{1})&&(2MASS-J\tablefootmark{3})& (Lyo\tablefootmark{2})& \\
 \hline
J0836 & M5.5 & M5.3 & M5.5 & 3058 & 2920 & 0.020 & 0.020 & 0.0190\\
RECX1 & K6 & K7.0 & K6.0 & 4205 & 4115 & 0.880 & 1.000 & 0.9800\\
J0838 & M5.25 & M5.0 & M5.5 & 3091 & 2920 & 0.034 & 0.035 & 0.0340\\
J0841 & M4.75 & M4.7 & M5.0 & 3161 & 2980 & 0.021 & 0.023 & 0.0210\\
RECX3 & M3.25 & M3.0 & M3.5 & 3379 & 3300 & 0.089 & 0.097 & 0.1000\\
RECX4 & M1.75 & M1.3 & M1.5 & 3596 & 3640 & 0.210 & 0.240 & 0.2000\\
RECX5 & M4 & M3.8 & M4.5 & 3270 & 3085 & 0.056 & 0.062 & 0.0670\\
RECX6 & M3 & M3.0 & M3.0 & 3415 & 3410 & 0.100 & 0.110 & 0.1000\\
RECX7 & K6 & K6.9 & K6.0 & 4205 & 4115 & 0.690 & 0.790 & 0.7000\\
J0843 & M3.25 & M3.4 & M4.0 & 3379 & 3190 & 0.074 & 0.083 & 0.0730\\
J0844 & M5.75 & M5.5 & M6.0 & 3024 & 2860 & 0.011 & 0.010 & 0.0078\\
RECX9 & M4.5 & M4.4 & M4.5 & 3198 & 3085 & 0.090 & 0.096 & 0.0950\\
RECX10 & M1 & M0.3 & M0.0 & 3705 & 3900 & 0.210 & 0.230 & 0.2100\\
RECX11 & K5.5 & K6.5 & K6.0 & 4278 & 4115 & 0.530 & 0.590 & 0.4600\\
RECX12 & M3.25 & M3.2 & M3.0 & 3379 & 3410 & 0.240 & 0.250 & 0.2300\\
 \hline 
\end{tabular} 
\tablebib{
 (1)~\citet{LuhmanSteeghs:2004aa};
 (2)~\citet{LyoLawson:2004aa};
 (3)~\citet{Cutriet-al.:2012aa,HerczegHillenbrand:2015aa}.
}
\label{tbl:stp_compare_lit}
\end{table*}			

We derive effective temperatures from spectral types using the conversion table 
proposed by \citet{HerczegHillenbrand:2014aa} \citep[see also][]{PecautMamajek:2013aa}, 
as estimated by comparing optical spectra of young templates to the
recent BT-Settl models of synthetic spectra \citep{AllardHomeier:2012aa}.

\smallskip
\noindent
Stellar luminosities are obtained from the continuum flux at
751{\us}nm and using the bolometric flux conversion table from
\citet{HerczegHillenbrand:2014aa}. For J0843 and J0844, the estimated flux in
the accretion continuum at 751{\us}nm (see Sect.~3.3) is subtracted before calculating $L_*$. The stellar radius is then
calculated from the Stefan Boltzmann law, $L_* = 4 \pi R_*^2 \sigma
T_{\mathrm{eff}}^4 $. The results are listed in Table
\ref{tbl:results}.  

\smallskip
\noindent
In Table \ref{tbl:stp_compare_lit} we compare the derived stellar
parameters with previous classifications of the $\eta$~Cha association 
\citep{LuhmanSteeghs:2004aa,LyoLawson:2004aa}. Spectral types agree
well within uncertainties. Consistent with the errors of the
classification, there is a small trend of spectral types in this work
being slightly earlier for early M stars and later for late M
stars. The effective temperatures are systematically colder for cold
objects, which is due to the different SpT-$T_{\mathrm{eff}}$ 
conversions used in this work. Use of the \citet{PecautMamajek:2013aa}
SpT-$T_{\mathrm{eff}}$ scale would have led to even lower temperatures
for these stars. 

\smallskip
\noindent
Measured stellar luminosities agree well within errors (see Table \ref{tbl:stp_compare_lit}) to both previous measurements \citep{LyoLawson:2004aa} and to luminosities inferred from $J$-Band (\citealt{Cutriet-al.:2012aa}; 
using bolometric corrections from either \citealt{PecautMamajek:2013aa} or
\citealt{HerczegHillenbrand:2015aa}).

\subsection{The Hertzsprung-Russell diagram}  \label{sec:stpa_thehrdiagram}
We use stellar isochrones from \citet{BaraffeHomeier:2015aa} to infer
stellar ages and masses. The isochrones are first mapped to a finer
grid in $T_{\rm eff}$. For a given grid-$T_{\rm eff}$ value, the
isochrones are interpolated onto a finer age grid. The adopted stellar
mass is given by the closest grid value in $T_{\rm eff}$ and $L_*$.  
  
\begin{figure}%[ht]		
 \centering			
 \includegraphics[width=1.\linewidth]{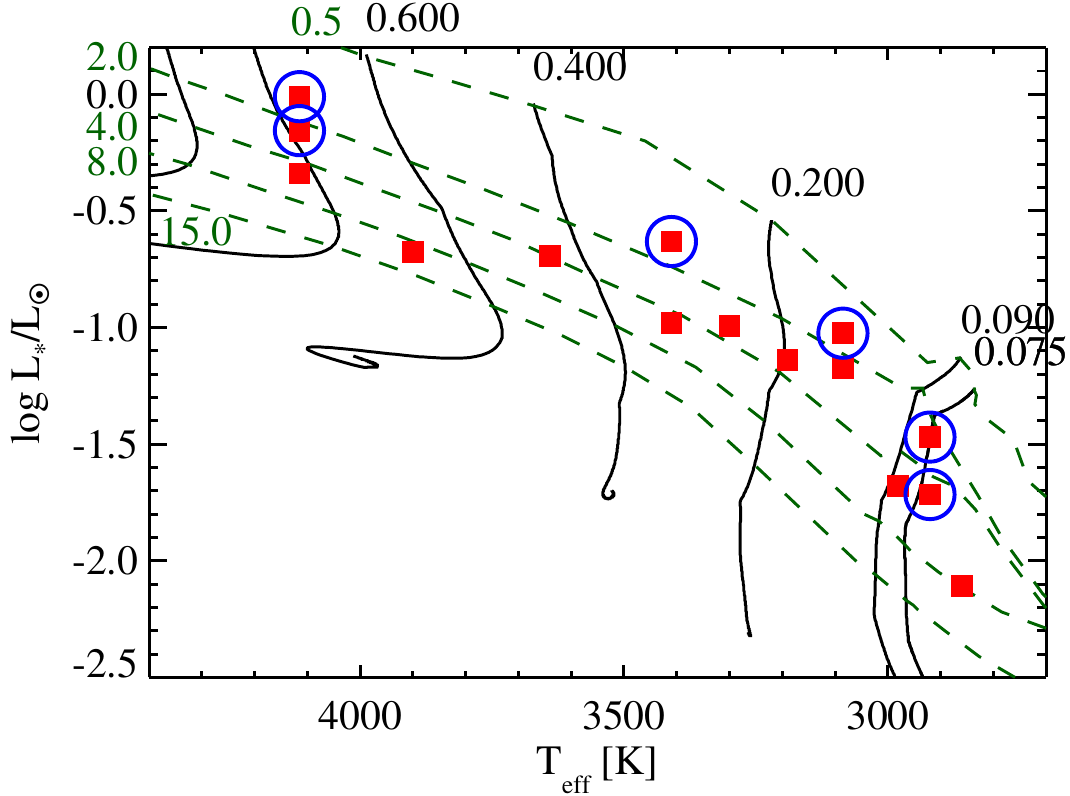}
 \caption{
  Hertzsprung-Russell diagram of the $\eta$ Cha association. Evolutionary tracks (mass in
  $M_\odot$) are shown as black solid lines and isochrones (age in Myr) as green dashed lines (both
  \citealt{BaraffeHomeier:2015aa}). $\eta$ Cha cluster stars are drawn as red squares. Binary stars are encircled in blue \citep[as in table 1 of][]{Sicilia-AguilarBouwman:2009aa}.
 }
 \label{fig:stp_hr_diag_baraffe15}
\end{figure}

\smallskip
\noindent
Fig.~\ref{fig:stp_hr_diag_baraffe15} shows the sample overplotted with
evolutionary tracks from \citet{BaraffeHomeier:2015aa}. A large
fraction of the sample (2/3) lies between the 1-Myr and 5-Myr
isochrones, approximately following the trend of the isochrones. The
known binaries are located above the single stars. The median 
age of single stars hotter than 3300{\us}K is 5{\us}Myr. This agrees
well with some of the previous results
\citep{LuhmanSteeghs:2004aa,HerczegHillenbrand:2015aa} and place
the $\eta$~Cha cluster at a significantly younger age than others
\citep[e.g., 11$\pm$3{\us}Myr,][]{BellMamajek:2015aa}. As a general
caveat, masses and ages inferred from Hertzsprung-Russell
diagrams may be misleading for low-mass stars when spectra are fit
with single-temperature photospheres
\citep[e.g.,][]{Gully-SantiagoHerczeg:2017aa}. 

\begin{table*}
 \caption{Derived stellar parameters in the $\eta$~Cha association}
 \centering	
\begin{tabular}{lcccccc} 
 \hline 
 \hline 
Object &$T_{\rm eff}$ &$ f_{751}$& $L_{\ast}$ &$R_{\ast}$ &$M_{\ast}$ &Infrared Class$^1$ \\
&[K]&$[1{\rm x }10^{-13}\,{\rm erg\,s^{-1}\,cm^{-2}\,\AA^{-1}}]$&[$L_{\odot}$]&[$R_{\odot}$]&[$M_{\odot}$]&\\
 \hline 
J0836&2920 &0.038&0.019&0.54&0.069&Class III\\
RECX1&4115 &3.100&0.980&1.90&0.750&Class III\\
J0838&2920 &0.066&0.034&0.72&0.075&Class III\\
J0841&2980 &0.043&0.021&0.54&0.086&TO/flat \\
RECX3&3300 &0.280&0.100&0.98&0.250&TO	  \\
RECX4&3640 &0.620&0.200&1.10&0.460&TO	  \\
RECX5&3085 &0.150&0.067&0.91&0.150&TO	  \\
RECX6&3410 &0.300&0.100&0.93&0.320&Class III\\
RECX7&4115 &2.200&0.700&1.60&0.780&Class III\\
J0843&3190 &0.180&0.073&0.88&0.200&Class II \\
J0844&2860 &0.014&0.008&0.36&0.052&flat	 \\
RECX9&3085 &0.220&0.095&1.10&0.150&TO	  \\
RECX10&3900&0.670&0.210&1.00&0.690&Class III\\
RECX11&4115&1.500&0.460&1.30&0.830&Class II \\
RECX12&3410&0.680&0.230&1.40&0.290&Class III\\
 \hline 
\end{tabular} 
 \tablebib{(1) Infrared disk classifications as in~\citet{Sicilia-AguilarBouwman:2009aa}.}
 \label{tbl:results}
\end{table*}

%%%%%%%%%%%%%%%%%%%%%%%%%%%%%%%%%%%%%%%%%%%%%%%%%%%%
% Mass Accretion Rates
%%%%%%%%%%%%%%%%%%%%%%%%%%%%%%%%%%%%%%%%%%%%%%%%%%%

\section{Mass accretion rate}
In this section, we measure accretion rates and upper
limits of our sample from UVB-arm spectra of X-Shooter. Section~\ref{sec:obsbalmerjump} discusses the
Balmer Jump as a diagnostic of excess continuum emission produced by
ongoing accretion. The excess flux is quantified by
modeling the observed spectrum, as described in
Sect.~\ref{subsec:modeling_uvexcess}, and used to derive the mass accretion rate. 
The model consists of non-accreting PMS spectra as proxies for photospheric emission of
young stars and the emission of a plane-parallel slab as proxy for the
radiation from regions heated in the accreting process. While the emission
from the plane-parallel hydrogen slab is not a physical model
\citep[for physical models see,
e.g.,][]{CalvetGullbring:1998aa,InglebyCalvet:2013aa}, the accretion
spectrum provides a reasonable fit to the continuum emission and a reasonable bolometric correction to account for emission
outside of the observed spectral range.
Individual objects are discussed in Sect.~\ref{sec:notes_individual_objects}.  

\subsection{The Balmer jump} \label{sec:obsbalmerjump}
\begin{figure}%[ht]
\centering
\includegraphics[width=1.\linewidth]{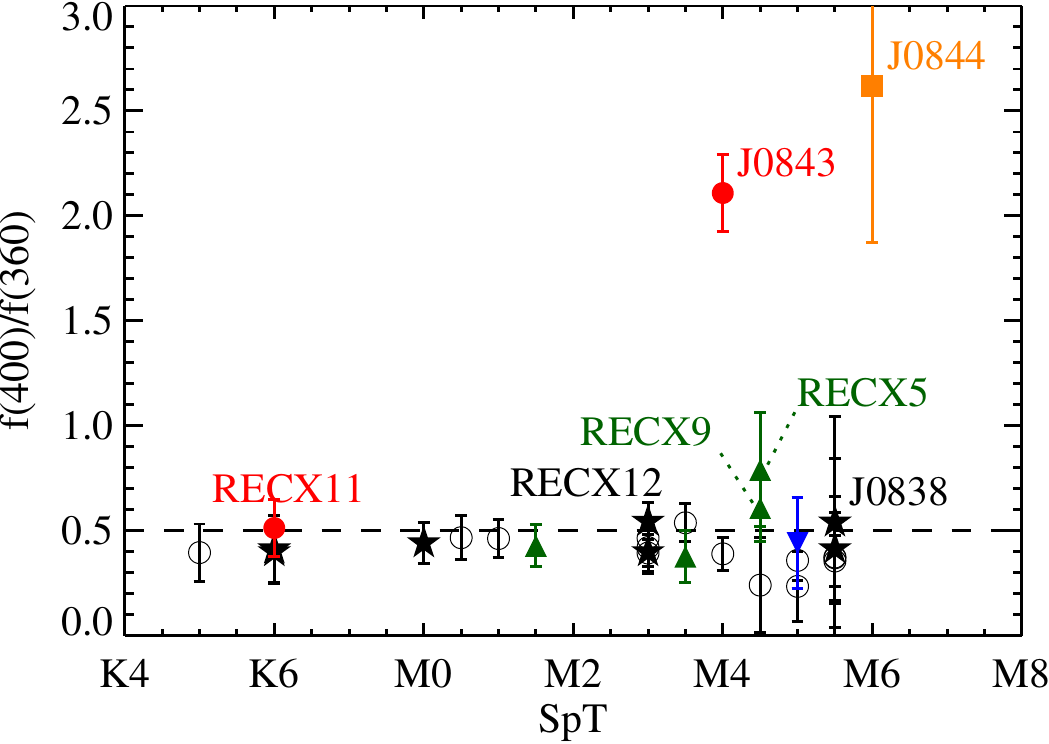}
\caption{
Observed Balmer jump of the $\eta$~Cha cluster members (filled
symbols) and non-accreting PMSs from \citet{ManaraTesti:2013aa},  
which were used as templates in this work (black empty circles). The
black dashed line highlights an observed Balmer jump of 0.5.  
The shape and color of the filled symbols highlight SED classifications
from \citet{Sicilia-AguilarBouwman:2009aa}: Black stars denote Class III objects,
red circles Class II objects, green triangles transitional objects, the blue, inverted 
triangle the transitional/flat object J0841, and the orange box the
flat source J0844.
}
\label{fig:plot_bj_temp_stars}%
\end{figure}

\begin{figure*}	%[ht]	
\centering			
\includegraphics[width=1.\linewidth]{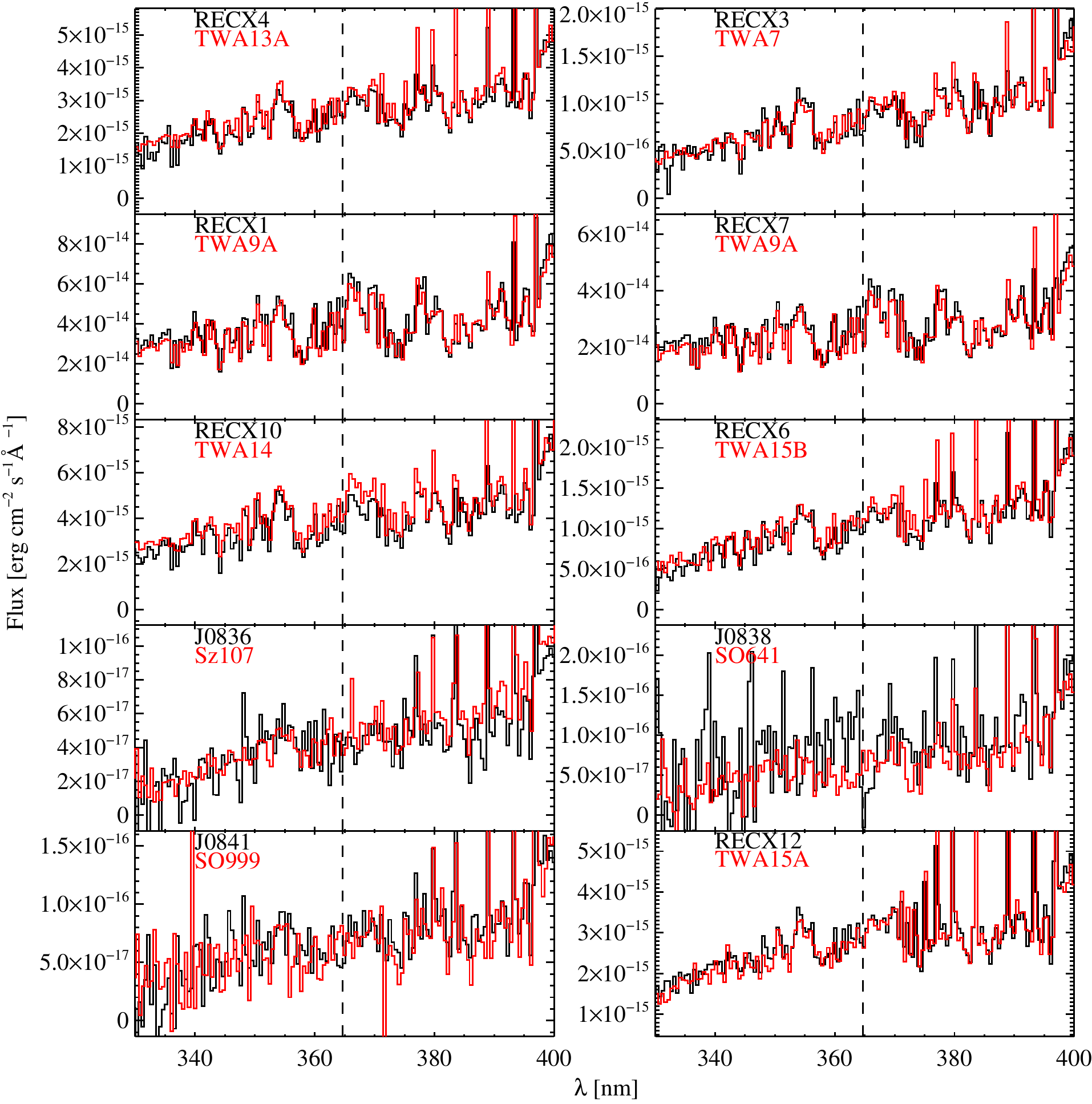}
\caption{
  Comparison of the UV spectrum for stars with an observed Balmer jump
  ratio typical for non-accreting stars (black), with template stars
  from \citet[][]{ManaraTesti:2013aa} of similar spectral type (red;
  template names as in \citealt[][]{ManaraTesti:2013aa}). The spectra
  are binned to 0.5{\us}nm resolution and the templates are scaled to
  the target at 450{\us}nm. The dashed vertical line shows the
  theoretical location of the Balmer jump.} 
  \label{fig:plot_nonaccreting}
\end{figure*}

\begin{figure*}%[ht]		
\centering			
\includegraphics[width=1.\linewidth]{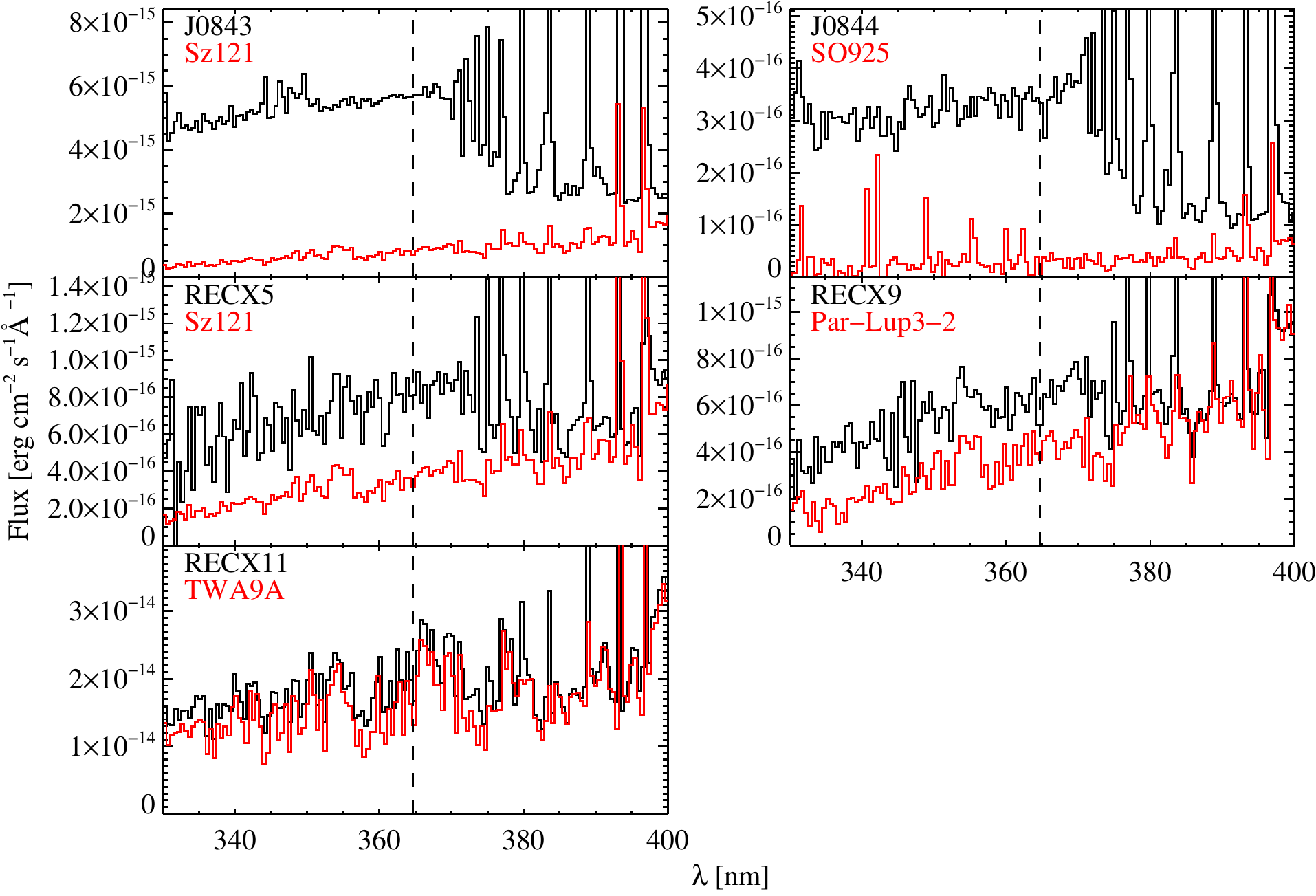}
\caption{
  As Fig. \ref{fig:plot_nonaccreting}. J0843, J0844 and RECX~5 and
  RECX~9 show clear indications of UV-excess. For RECX~11, a slight
  excess is seen. 
}
\label{fig:plot_maybeaccreting}
\end{figure*}

The presence of excess Balmer-continuum emission produced by accretion
is quantified here by measuring Balmer Jump, defined as the flux ratio 
$f(360{\us}\textrm{nm})/f(400{\us}\textrm{nm})${}\footnote{ The
  Balmer Jump measured here includes the photosphere and is not a
  photosphere-subtracted measurement of the accretion continuum.}. The X-Shooter UVB
spectra are shown in Figures~\ref{fig:plot_nonaccreting}~and~\ref{fig:plot_maybeaccreting} along with template, non-accreting, PMS
stars \citep{ManaraTesti:2013aa,StelzerFrasca:2013aa}, with results on
the Balmer jump shown in Fig.~\ref{fig:plot_bj_temp_stars}.
The dashed line ($f(360{\us}\textrm{ nm})/f(400{\us}\textrm{nm})$ =
0.5) indicates the threshold between 
accreting and non-accreting sources \citep{HerczegCruz:2009aa},
consistent with the non-accreting templates of \citet{ManaraTesti:2013aa}.

\smallskip
\noindent
Of the $\eta$~Cha cluster members, seven sources show a Balmer Jump 
above this threshold.  Three sources have strong UV-excess emission
(RECX~5 (Balmer Jump $\sim$ 0.8),  
J0843 (2.1) and J0844 (2.6)), while four have a moderate excess (between
0.5$-$0.6: RECX~11, RECX~12, RECX~9 and J0838). In the case of J0838,
the continuum level at 360~nm is enhanced with respect to the
template, while at wavelengths below 355~nm the spectrum agrees with
the template within the noise (Fig.~\ref{fig:plot_nonaccreting}). The
enhanced Balmer Jump of this star is therefore not considered as an
indicator of accretion. The UV-excess in RECX~12 is attributed to
chromospheric activity, as the H$\alpha$ line is narrow and has an equivalent width which is lower
than expected for mass accretion (see Sect.~\ref{sec:lineemission}),
and has a UV spectrum consistent with the non-accreting template TWA
15A (Fig.~\ref{fig:plot_nonaccreting}). RECX~12 is
discussed in more detail in Sect.~\ref{sec:lineemission}. The
UV-excess is clearly visible in the spectra of RECX~9 and RECX~11.

\subsection{Method of Fitting the UV-excess} \label{subsec:modeling_uvexcess}
The accretion continuum, modeled from a \mbox{pure-hydrogen}
isothermal slab, is added to the
photospheric template (see Sect.~\ref{sec:stellarparametersgridcomparison}) to best fit the UV spectrum of the accreting
star.   The grid of slab models \citep{ValentiBasri:1993aa} spans a parameter space in
density and temperature of the hydrogen gas (Table~\ref{tbl:parameterrange_slab}), where the length of the
slab is fixed to $l = 2 \textrm{ x } 10^7${\us}cm
\citep{HerczegCruz:2009aa}.   
The bolometric correction is
calculated by summing the accretion continuum luminosity over all wavelengths.

\smallskip
\noindent
Best-fit parameters are calculated by minimizing a
$\chi^2$-like function over multiple wavelength bands in the UVB
arm of X-Shooter, with wavelengths selected
to exclude sharp emission and absorption lines.
The $\chi^2$-like minimization function is defined as $\chi^2_{\mathrm{like}}=\sum_{\mathrm{bands}}\left({\frac{\overline{f_{\mathrm{obs}}}-\overline{f_{\mathrm{model}}}}{\sigma(f_{\mathrm{obs}})}}\right)$, where $\overline{f_{\mathrm{obs}}}$ and
$\overline{f_{\mathrm{model}}}$ denote the median of each band in the
observed spectrum and model, and $\sigma(f_{\mathrm{obs}})$
gives the standard deviation in each band in the observed spectrum. As also described in \citet{ManaraBeccari:2013aa}, this does not
represent a true $\chi^2$ distribution, as the noise in the template spectrum
is neglected. The models are visually compared to confirm that
the \mbox{best-fit} model matches the observed spectrum.
The uncertainties are estimated from the range of UV-excess fluxes
spanned by models that yield a
similar $\chi^2$-like value as the best-fit model and match the
spectrum visually (details are described in
Sect.~\ref{sec:notes_individual_objects}).

\begin{table}%[ht]
\caption{Parameter ranges of the hydrogen slab model}
\centering	
\begin{tabular}{l|r|r|r} 
 \hline 
 \hline 
  &Min. & Max. & Step size\\
 \hline 
	$T_{\rm eff}{\rm[K]}$	&6000	&11000	&500\\
	${\rm log(n_H[cm^{-3}])}$	&	12	&	15	& 0.01\\
 \hline 
\end{tabular} 
\label{tbl:parameterrange_slab}
\end{table}		

\smallskip
\noindent
The measured UV-excess flux is  converted to accretion luminosity
($L_{\rm acc}$).  The mass accretion rates ($\dot{M}$) are then calculated by 
\begin{equation}	
 L_{\mathrm{acc}}=\frac{GM_*\dot{M} }{R_*} \cdot \left( 1-\frac{R_*}{R_{\mathrm{trunc}}}\right),
 \label{eq:macc}
\end{equation}
 under
 the assumption that gas accretes from the disk truncation radius $R_{\mathrm{trunc}}$ in
 free-fall onto the star, where $R_{\rm trunc}=5\,R_*$ is adopted for consistency with pre-existing measurements
\citep[see review by][]{HartmannHerczeg:2016aa}.

\subsection{Description of fits and uncertainties} \label{sec:notes_individual_objects}
Figures \ref{fig:plot_w_models_s015} -
\ref{fig:plot_w_models_s042} show the \mbox{best-fit} model for each object.
$L_{\rm acc}$ and $\dot{M}$ are listed in Table~\ref{tbl:comp_acc_measurements}.   

\smallskip
\noindent
Uncertainties in the accretion rates are introduced by uncertainties
in the stellar parameters, the distance, and in the individual fits
\citep[see, e.g.,][]{HerczegHillenbrand:2008aa,ManaraBeccari:2013aa}. 
For our sample, uncertainties in $\dot{M}$ are $\sim$0.3~dex.
Weak accretors, such as RECX~11, have uncertainties of $\sim$0.5~dex
because the excess is fainter than the photosphere, thereby increasing the uncertainty in
bolometric correction.  

\begin{figure*}		
\centering			
\includegraphics[width=0.99\linewidth]{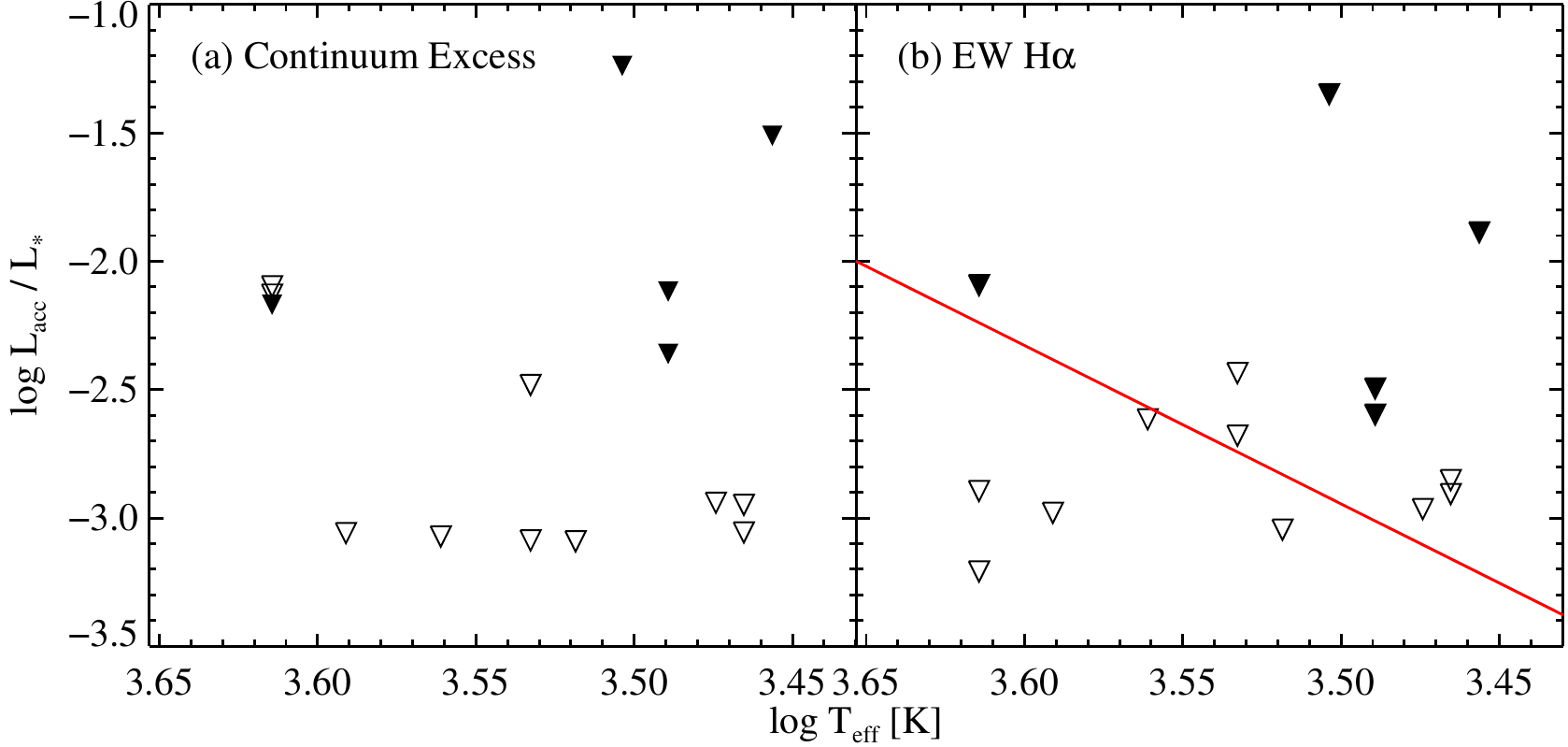}
\caption{Accretion luminosity (in units of $L_*$) vs. $T_{\rm eff}$ for accreting PMS (filled triangles) and upper limits for non-accreting PMS (empty triangles). In panel (a), $L_{\rm acc}$ is derived from continuum excess radiation and in panel (b) from the H$\alpha$ equivalent width. The red line in panel (b) shows the chromospheric contribution from hydrogen emission lines to $L_{\rm acc}$ \citep{ManaraTesti:2013aa}. }
\label{fig:plot_lmacc_teff}
\end{figure*}

\smallskip
\noindent
Upper limits on $\dot{M}$ are difficult to estimate and depend on the ability to
distinguish any excess emission from the underlying photospheric
spectrum and on an uncertain bolometric correction.
In order to estimate upper limits of $\dot{M}$ for non-accreting
objects, we perform a similar model fit as described above.
The upper limits are compared to the detections in Fig.~\ref{fig:plot_lmacc_teff}a in terms of the ratio of 
$L_{\rm acc}/L_*$. For K stars, the upper limit of 
$L_{\rm acc}/L_*$ for non-accreting stars is similar to $L_{\rm acc}/L_*$ of RECX~11. 
For M stars, the upper limits are rather uniform at 
${\rm log}\,L_{\rm acc}/L_* \approx -3.0$ \citep[see also][]{ManaraTesti:2013aa}, with the 
exception of RECX~12 (${\rm log}\,L_{\rm acc}/L_* \approx -2.7$). 

\smallskip
\noindent
Comments on each fit to the spectra of accreting objects are provided below.

\smallskip
\noindent
{\it J0843 and J0844:}  Due to the pronounced observed Balmer jump and strong excess
in Balmer and Paschen continuum, the slab model parameters and
therefore the $L_{\rm acc}$ are well
constrained.  Models using templates within 0.5
spectral subclasses yield accretion rates that agree within 10\% of
these values. 

\smallskip
\noindent
{\it RECX~5:} The best-fit $\dot{M}$ is calculated using the
M4 template Sz 121.  Model results using the M3 template TWA 9B and or
the M5 template Par-Lup3-2 also reproduce the UVB spectrum of RECX 5,
yield accretion rates that differ by 30\% from the adopted best-fit.

\smallskip
\noindent
{\it RECX~9:}  The best-fit $\dot{M}$ is calculated from the
M4.5 template SO797.  Model results
using templates the M5 templates Par-Lup3-2 and SO641 yield accretion
rates that differ by 45\%.

\smallskip
\noindent
{\it RECX~11:} $\dot{M}$ of RECX~11 is uncertain by a factor
of 2.5 because the UV excess is very small, so the slab model fit is
not well constrained.

%%%%%%%%%%%%%%%%%%%%%%
%LINE EMISSION
%%%%%%%%%%%%%%%%%%%%%%
\section{Line emission} \label{sec:lineemission}

Accretion and related processes produce emission in many detectable
lines, a subset of which are analyzed
here: The H$\alpha$ and H$\beta$ transitions, the
\ion{He}{i}${~\lambda 402.6}$, ${{\rm HeI}~\lambda 447.1}$,
\ion{He}{i}${~\lambda 501.5}$, \ion{He}{i}${~\lambda 587.5}$,
\ion{He}{i}${~\lambda 667.8}$, \ion{He}{i}${~\lambda 706.5}$
transitions and blended \ion{He}{i} and \ion{Fe}{i} lines at 492.2 nm
\citep[$\ion{He}{i}\ion{Fe}{i}\lambda 492.2$,][]{AlcalaNatta:2014aa},
and the [\ion{O}{i}]${~\lambda 557.7}$, [\ion{O}{i}]${~\lambda 630.0}
$, [\ion{O}{i}]${~\lambda 636.3}$ forbidden transitions. The flux and
equivalent widths of the detected lines is given in
Tables~\ref{tbl:comp_acc_measurements} 
and \ref{tbl:emission_1} to \ref{tbl:emission_4}.

\smallskip 
\noindent
Figures~\ref{fig:halphaline} and~\ref{fig:hbetaline} show the
H$\alpha$ and H$\beta$ profiles. The detection
rate of the two transitions is 100\% for H$\alpha$ and $\sim$90\% for
H$\beta$ throughout the full sample. 
For J0843, J0844, RECX~5, RECX~9 and RECX~11, the H$\alpha$ line profiles are very broad ($\Delta v >
200{\us}$km~s$^{-1}$) and slightly asymmetric, consistent with
accretion. All other sources have narrow, symmetric H$\alpha$ profiles
with equivalent widths that are
below the mass accretion threshold \citep{WhiteBasri:2003aa},
consistent with a chromospheric origin of the line emission. 

\smallskip
\noindent
Transitions of \ion{He}{i} are detected in J0844, J0843, RECX~5, RECX~9
and RECX~12 (see Tables \ref{tbl:emission_1} to
\ref{tbl:emission_4}). Emission in \ion{He}{i}${~\lambda 402.6}$ is
weakly detected in J0841. J0836 shows weak emission in the ${{\rm
    HeI}~\lambda 447.1}$ line. The emission lines for  RECX~12 are in
all cases weaker than for the accreting stars with detections in
\ion{He}{i}. To note special line profiles: J0843 shows a broad, blue
shifted component at lower flux than the central, narrow peak in the
${{\rm HeI}~\lambda 447.1}$ transition and broad wings in the
\ion{He}{i}${~\lambda 587.5}$ transition. The \ion{He}{i}${~\lambda
  501.5}$ line and \ion{He}{i}+\ion{Fe}{i} blend at $\lambda$~492.2 are blended
with \ion{Fe}{ii} emission at slightly higher wavelengths for J0844
and J0843. \ion{Fe}{ii} emission has also been found in heavily veiled
objects, including some outbursts
\citep[e.g.,][]{HessmanEisloeffel:1991aa,HamannPersson:1992aa,van-den-AnckerBlondel:2004aa,Fedelevan-den-Ancker:2007aa,GahmWalter:2008aa}.  

\begin{figure}%[ht]		
\centering			
\includegraphics[width=1.\linewidth]{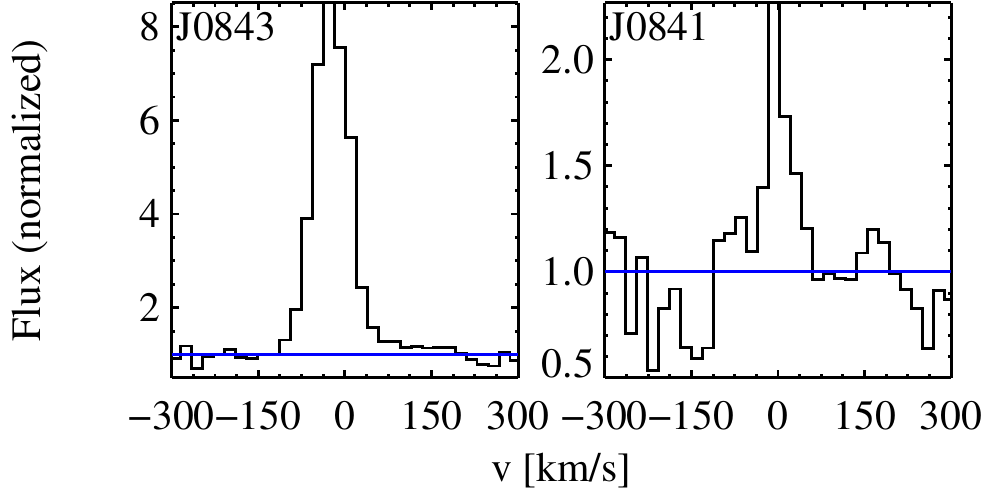}
\caption{Continuum normalized [\ion{O}{i}] ${~\lambda 630.0}${\us}nm
  line profiles in J0843 and J0841. In J0843, the line is
  blue-shifted. The continuum level is indicated by a blue line.} 
\label{fig:plot_oI6300}
\end{figure}

\smallskip
\noindent
The forbidden [\ion{O}{i}]${~\lambda 630.0}${\us}nm line, a diagnostic
of winds \citep[e.g.][]{HartiganEdwards:1995aa,SimonPascucci:2016aa}, is detected
from J0843 and J0841, and marginally (<3$\sigma$) from J0844 and
RECX~5. To characterize velocity shifts of the line center, the
wavelength solution was re-calibrated at the photospheric
\ion{Li}{I}~$\lambda$670.8 nm line\footnote{The radial velocity of the two
detected objects are measured at \mbox{11--13{\us}km{\us}s$^{-1}$} and are
similar to the mean radial velocity of the $\eta$~Cha cluster
(14$\pm$1{\us}km{\us}s$^{-1}$; e.g.,
\citealt{Lopez-MartiJimenez-Esteban:2013aa}).}.  
The [\ion{O}{i}]${~\lambda 630.0}$\,nm line is blueshifted by
24{\us}km{\us}s$^{-1}$ for J0843, and not detectably shifted in J0841
(Fig. \ref{fig:plot_oI6300}). Additionally, the [\ion{O}{i}]${~\lambda
  557.7}$ transition is detected from J0843, J0844, RECX~5 and RECX~9,
while the [\ion{O}{i}]${~\lambda 636.3}$ line is only detected from J0843.  

\subsection{Notes on individual objects} \label{sec:lines_notes_on_individual_objects}
RECX~12 and J0838 both show a slightly enhanced Balmer
jump (Sect.~\ref{sec:obsbalmerjump}), along with an equivalent width of the
H$\alpha$ line which is consistent with chromospheric emission. We therefore
categorize both stars as non-accreting. However, the status of RECX~12 as a non-accretor is uncertain.  The
H$\alpha$ line has weak line wings that are 
at levels of smaller than 10\% of the peak flux and are asymmetric towards
the blue side of the line. The H$\alpha$ 10\% width (200$\pm$20{\us}km{\us}s${}^{-1}$) lies at the threshold for accretion
\citep[e.g.,][]{WhiteBasri:2003aa,JayawardhanaMohanty:2003aa,NattaTesti:2004aa}. 
Three \ion{He}{i} transitions are detected above 3$\sigma$ for RECX~12, but their
equivalent widths are lower than any of the clearly accreting objects in
the sample and are consistent with strong chromospheric emission. The
upper limit on $L_{\rm acc}/L_*$, as derived from continuum excess and H$\alpha$ emission, is higher 
than for non-accreting objects of similar spectral types in the $\eta$
Cha cluster (see Fig.~\ref{fig:plot_lmacc_teff}a, b). If the emission
was interpreted as accretion, then $\dot{M}$ determined
from the H$\alpha$ transition and the continuum emission would agree
well with each other, while the H$\beta$ and \ion{He}{i} emission lines would yield
accretion rates that would be 2$-$3 times larger. $\dot{M}$ would
therefore be between $1.5-5.0\times10^{-10} {\us}
{\rm M}_{\odot} / {\rm yr}$.  However, since some non-accreting young stars
share these characteristics (e.g., the PMS TWA15A in
\citealt{ManaraTesti:2013aa}), we interpret this borderline case as
more likely a source with strong chromospheric emission. 

\smallskip
\noindent
The detection of [\ion{O}{i}] emission in J0841 is unusual, since
accretion is not detected in either H$\alpha$ or in the
Balmer continuum.  However, a disk is detected from excess infrared
emission \citep[e.g.,][]{Sicilia-AguilarBouwman:2009aa}.
The \ion{He}{i}~$\lambda$402.6 emission in this object is barely
detected. Emission of the [\ion{O}{i}] line can originate from disk
winds, which can be related to the accretion process (e.g.,
\citealt{HartiganEdwards:1995aa}; Nisini et al., submitted). Mass accretion rates 
between $1.2-14\times10^{-11} {\us} {\rm
  M}_{\odot} / {\rm yr}$ would be required to produce the luminosity 
of the [\ion{O}{i}] line ($L_{\rm line}$), assuming a \mbox{$L_{\rm acc}$-$L_{\rm line}$ relation}
from \citet{NattaTesti:2014aa}. The upper limit on $\dot{M}$ 
from continuum excess emission of $\dot{M} \approx
0.6\times10^{-11} {\us} {\rm M}_{\odot} / {\rm yr}$ falls slightly
below this range. Hence, the origin of [\ion{O}{i}] emission in J0841
is unclear.  

\smallskip
\noindent
H$\alpha$ emission of RECX~7 shows a double-peaked profile and has
been classified as a spectroscopic binary in the literature
\citep{MamajekLawson:1999ab,LyoLawson:2003aa}. The equivalent width of
each component is in agreement with chromospheric emission. The peaks
are separated by $\Delta v \approx$
185{\us}km{\us}s$^{-1}$.  Indication of H$\beta$ emission may be seen in
the low velocity component (see lower right-most panel of
Fig.~\ref{fig:hbetaline}) but is confused by absorption lines and is
not considered significant.

%TABLE WITH ALL RESULTS
\begin{table*}%[ht]
\caption{Mass accretion properties in the $\eta$~Cha association}
\centering	
\begin{tabular}{l|r|r|r|r|r|r|r|r} 
 \hline 
 \hline 
 Object &$EW_{{\rm H\alpha}~\lambda 656.3} $ & $L_{{\rm acc,H} \alpha} $ &$\dot{M}_{{\rm H} \alpha} $ &$EW_{{\rm H\beta}~\lambda 486.1} $&$L_{{\rm acc,H} \beta} $ &$\dot{M}_{{\rm H} \beta} $ &$L_{{\rm acc, cont} } $ &$\dot{M}_{{\rm cont}} $\\
  &[\AA]& [L$_\odot$/yr] &[M$_\odot$/yr] &[\AA] &[L$_\odot$/yr] &[M$_\odot$/yr]&[L$_\odot$/yr] &[M$_\odot$/yr] \\
 \hline 
J0836 &$-$13.4$\pm$0.7 & <2.4e-05& <7.5e-12&$-$11.1$\pm$0.4 & <2.4e-05 & <7.6e-12 &<1.7e-05	&<5.3e-12 \\
RECX1 &$-$1.0$\pm$0.1  & <9.0e-04& <9.4e-11&<0.2	    & <1.7e-04 & <1.8e-11 &<7.2e-03	&<7.5e-10 \\
J0838 &$-$12.0$\pm$0.8 & <4.6e-05& <1.8e-11&$-$10.4$\pm$0.4 & <5.2e-05 & <2.0e-11 &<3.8e-05	&<1.5e-11 \\
J0841 &$-$9.5$\pm$0.6  & <2.1e-05& <5.3e-12&$-$8.3$\pm$0.3  & <2.7e-05 & <6.9e-12 &<2.4e-05	&<6.1e-12 \\
RECX3 &$-$2.7$\pm$0.3  & <8.7e-05& <1.4e-11&$-$1.9$\pm$0.1  & <1.1e-04 & <1.7e-11 &<8.2e-05	&<1.3e-11 \\
RECX4 &$-$4.2$\pm$0.3  & <4.9e-04& <4.9e-11&$-$2.7$\pm$0.2  & <7.3e-04 & <7.2e-11 &<1.7e-04	&<1.7e-11 \\
RECX5 &$-$13.5$\pm$0.7 &  2.1e-04&  5.2e-11&$-$14.7$\pm$0.4 &  4.5e-04 &  1.1e-10 & 5.1e-04	& 1.3e-10\\
RECX6 &$-$5.3$\pm$0.2  & <2.2e-04& <2.5e-11&$-$4.0$\pm$0.1  & <3.4e-04 & <4.0e-11 &<8.6e-05	&<1.0e-11 \\
RECX7 &$-$0.8$\pm$0.1  & <4.3e-04& <3.6e-11&<0.2	    & <9.5e-05 & <8.1e-12 &<5.6e-03	&<4.7e-10 \\
J0843 &$-$94.5$\pm$2.6 &  3.2e-03&  5.8e-10&$-$42.6$\pm$0.7 &  3.3e-03 &  5.9e-10 & 4.2e-03	& 7.6e-10\\
J0844 &$-$102.2$\pm$5.5&  9.8e-05&  2.7e-11&$-$82.1$\pm$2.0 &  2.0e-04 &  5.5e-11 & 2.4e-04	& 6.6e-11\\
RECX9 &$-$12.6$\pm$0.5 &  2.4e-04&  6.8e-11&$-$10.7$\pm$0.3 &  3.1e-04 &  9.0e-11 & 4.1e-04	& 1.2e-10\\
RECX10 &$-$1.7$\pm$0.2 & <2.2e-04& <1.3e-11&$-$1.2$\pm$0.2  & <4.2e-04 & <2.5e-11 &<1.8e-04	&<1.1e-11 \\
RECX11 &$-$8.3$\pm$0.3 &  3.9e-03&  2.5e-10&$-$2.7$\pm$0.2  &  5.0e-03 &  3.2e-10 & 3.1e-03	& 2.0e-10\\
RECX12 &$-$8.4$\pm$0.2 & <8.6e-04& <1.6e-10&$-$8.2$\pm$0.2  & <1.9e-03 & <3.6e-10 &<7.7e-04	&<1.5e-10 \\
 \hline 
\end{tabular} 
\tablefoot{Columns 1--7 list equivalent width (EW), accretion
  luminosity ($L_{\rm acc}$; using conversions from \citet{AlcalaNatta:2014aa}) and
  mass accretion rate ($\dot{M}$) for H$\alpha$ and H$\beta$, respectively. The
  last two columns list $L_{\rm acc}$ and $\dot{M}$ determined from UV~excess. } 
\label{tbl:comp_acc_measurements}
\end{table*}	

\subsection{Mass accretion rates from Hydrogen emission lines}
Table \ref{tbl:comp_acc_measurements} lists $L_{\rm acc}$ and $\dot{M}$ derived from the line luminosities of H$\alpha$ and H$\beta$ using empirical $L_{\rm acc}$-$L_{\rm line}$ relations \citep{AlcalaNatta:2014aa}. The ratio $L_{\rm acc, H\alpha}/L_*$ is shown in Fig.~\ref{fig:plot_lmacc_teff}b.  

\smallskip
\noindent
For J0843, the accretion rates derived from different tracers agree within 10\%.  
For J0844, RECX~5 and RECX~9, $\dot{M}_{\rm H\alpha}$ is lower by about a factor of two, 
however $\dot{M}_{\rm H\beta}$ is closer to and within errors of $\dot{M}$ obtained from the continuum excess. 
A possible reason for this discrepancy could be that the H$\alpha$ line becomes easily optically thick and may be influenced by outflows \citep[e.g.,][]{AlcalaNatta:2014aa}.
Accretion rates, which are derived from $L_{\rm line, H\alpha}$, agree closely with the accretion rates
determined by direct modeling of the H$\alpha$ line profile
\citep{LawsonLyo:2004aa}. 

\smallskip
\noindent
The H$\alpha$ equivalent width of RECX~11 is 
significantly higher than for the other K stars in this sample (RECX~1 and RECX~7) and falls 
in the range which is typically found for accreting stars 
at this spectral type \citep{WhiteBasri:2003aa}, despite weak
continuum emission (see Fig.~\ref{fig:plot_lmacc_teff}).  The
H$\alpha$ line of RECX~11 is also affected by red-shifted absorption from the
accretion flow \citep[see also][]{InglebyCalvet:2011aa}.
$\dot{M}_{\rm H\alpha}$ and $\dot{M}_{\rm H\beta}$ are higher than 
but within the uncertainties of $\dot{M}$ determined from UV-excess.

\begin{figure*} \centering \includegraphics[width=.7\linewidth]{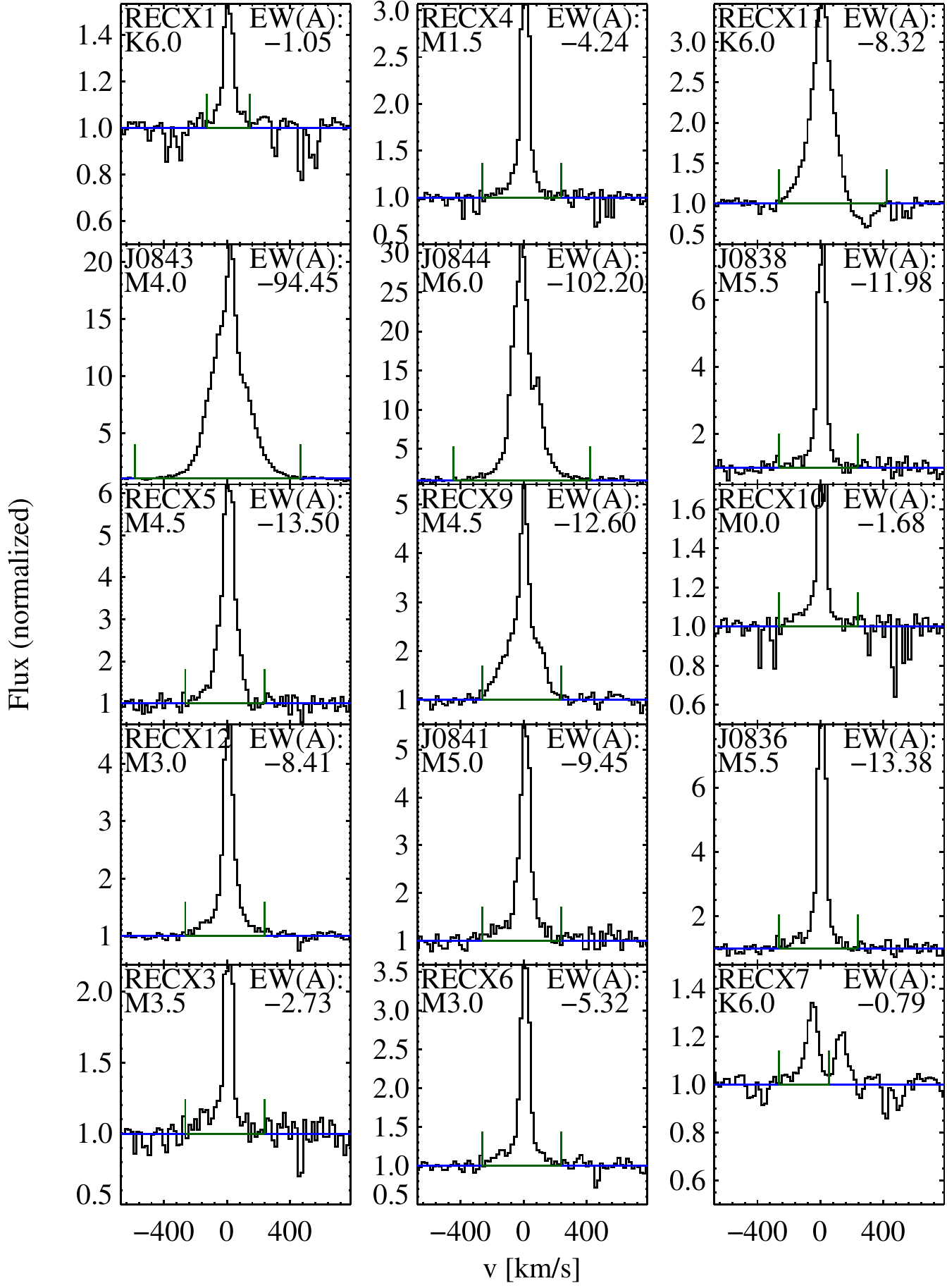} 
\caption{Continuum normalized H$\alpha$ line profiles in the
  $\eta$~Cha association. Green lines denote the integration
  boundaries for the determination of line flux and equivalent
  width. The continuum level is indicated by a blue
  line. Each panel shows the object's name, its spectral type and the measured equivalent width (EW).} \label{fig:halphaline} \end{figure*} 
\begin{figure*}
  \centering \includegraphics[width=.7\linewidth]{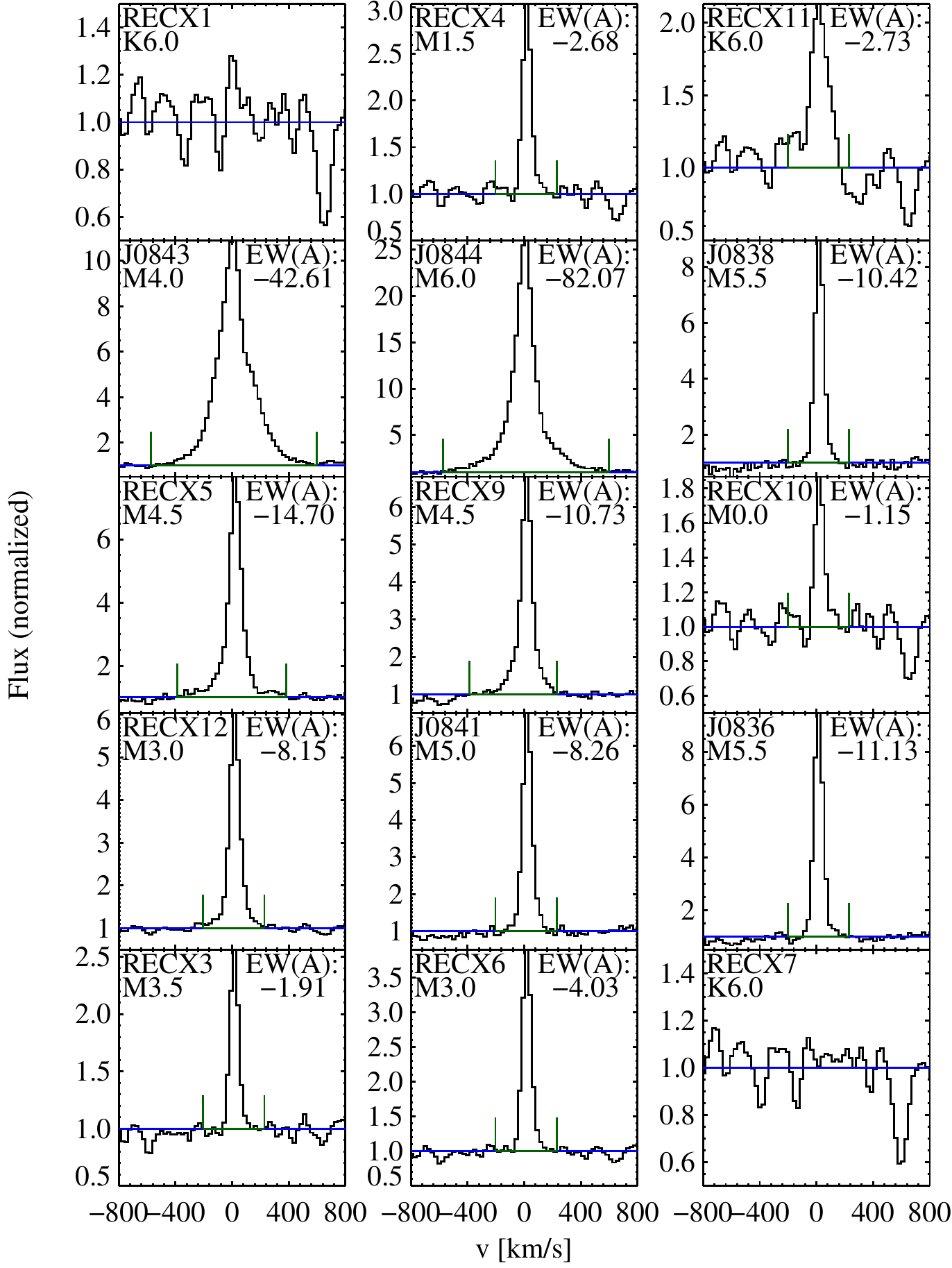} \caption{Line
    profiles of the H$\beta$ ${\lambda 486.13}$nm transition. Symbols
    and colors as in
    Fig. \ref{fig:halphaline}.} \label{fig:hbetaline} \end{figure*} 

%%%%%%%%%%%%%%%%%%%%%%%%%%%%%%%%%%%%%%%%%%%%%%%%%%%
% Discussion
%%%%%%%%%%%%%%%%%%%%%%%%%%%%%%%%%%%%%%%%%%%%%%%%%%%
\section{Discussion} \label{sec:discussion}
%%%%%%%%%%%%%%%%%%%%%%%%%%%%%%%%%%%%%%%%%%%%%%%%%%%%%
\subsection{Detectability of mass accretion}
H$\alpha$ emission is detected in all sources, generated by either
the accretion flow or chromospheric activity. As seen in
Fig.~\ref{fig:plot_lmacc_teff}b, the $L_{\rm acc}/L_*$ ratio of accreting stars is higher than the upper limits inferred for non-accreting PMS. The
upper limits on $L_{\rm acc, H\alpha}/L_*$ are consistent with the threshold for "accretion noise" inferred by \citet{ManaraTesti:2013aa}.  

\smallskip
\noindent
Detection limits for mass accretion rates from UV-excess depend on the spectral type of the
star. For M-stars, the detection limits are at $\dot{M}\approx 1 \times 10^{-11} {\us} {\rm M}_{\odot} / {\rm yr}$ (Table~\ref{tbl:comp_acc_measurements}).
For the K-stars RECX~7 and RECX~1 (${\rm log}T_{\rm eff}>3.6$), the upper limits of \mbox{$\dot{M}\approx 4-8 \times 10^{-10} {\us} {\rm M}_{\odot} / {\rm yr}$} are higher than $\dot{M}$ of the accretor RECX~11. Unlike for M-stars, these upper limits do not reflect the detection limit but the large uncertainty in fitting the weak \mbox{UV-excess} emission in the warmer K-stars (see Sect.~\ref{sec:notes_individual_objects}). 

%%%%%%%%%%%%%%%%%%%%%%%%%%%%%%%%%%%%%%%%%%%%%%%%%%%%%

\subsection{Variability}
The accretion rate has been previously measured on 4 of the 5
accretors in $\eta$~Cha identified here.  Our accretion rates obtained
from excess Balmer continuum emission are
consistent with past measurements using similar approaches but in
several cases differ when compared with H$\alpha$-based measurements.

\smallskip
\noindent
{\it J0843:} Our accretion rate is similar to the UV-excess accretion
rate $8 \times 10^{-10} {\us} {\rm M}_{\odot} / {\rm yr}$ 
measured by \citet{InglebyCalvet:2013aa} and is 1.25 times
lower than the rate of $1 \times 10^{-9} {\us} {\rm M}_{\odot} / {\rm
  yr}$ measured from modeling the H$\alpha$ line by 
\citet{LawsonLyo:2004aa}. All three measurements agree within errors. 

\smallskip
\noindent
{\it RECX 9:} Our accretion rate is larger than previous measurements by a
factor of three (\citealt{LawsonLyo:2004aa}; 
$4 \times 10^{-11}{\us}{\rm M}_{\odot} / {\rm yr}$). The H$\alpha$
line profile and equivalent width from Fig.~1 in 
\citet{LawsonLyo:2004aa} are similar to those measured in this work
(central panel in Fig. \ref{fig:halphaline}), so differences are
likely the result of methodology.

\smallskip
\noindent
{\it RECX 5:}  As with RECX 9, $\dot{M}$ of RECX 5
measured here is two times larger than that obtained from H$\alpha$ modeling \citep[$5
\times 10^{-11} {\us} {\rm M}_{\odot} / {\rm
  yr}$;][]{LawsonLyo:2004aa}.  This difference is unlikely due to
variability because the Lawson et al.~spectrum had a much stronger and
broader H$\alpha$ emission: an equivalent
width of of $-35$ \AA, a 10\% width of 330 km s$^{-1}$, and a FWHM of 160  km s$^{-1}$ in the Lawson et
al.~spectrum, compared to $-14$ \AA, 220 km s$^{-1}$, and 105  km
s$^{-1}$ in our spectrum.  The variability in H$\alpha$ emission from
RECX 5 had
been found previously \citep{JayawardhanaCoffey:2006aa,MurphyLawson:2011aa}.

\smallskip
\noindent
{\it RECX 11:} $\dot{M}$ from this work is similar to the
value found by previous measurements of UV-excess ($1.7
\times 10^{-10} {\us} {\rm M}_{\odot} / {\rm yr}$;
\citealt{InglebyCalvet:2011aa,InglebyCalvet:2013aa}).  However, these
accretion rates are both five times higher than
that determined from H$\alpha$ line profile modeling \citep[$4 \times
10^{-11} {\us} {\rm M}_{\odot} / {\rm yr}$;][]{LawsonLyo:2004aa}.  The
H$\alpha$ line is stronger in our observations ($-8.3$ \AA\ equivalent
width) than in the Lawson et al. spectrum ($-3$ \AA), so the different
accretion rates may result from methodology, variability, or
uncertainty in accretion measured from the weak UV excess.

\smallskip
\noindent
{\it J0844:}  No previous direct measurement of mass accretion onto
J0844 exists in the literature. The equivalent width of the H$\alpha$ line in the
X-Shooter spectrum is nearly twice as high as previously reported ($-$57.8{\us}\AA;
\citealt{SongZuckerman:2004aa}), with a range in variability
consistent with that seen from other sources \citep{CostiganScholz:2012aa}. 
$\dot{M}$ of J0844 is located at the high end of the
scatter in $\dot{M}$ of accreting low-mass stars (Fig.~\ref{fig:logm_logmacc}; see Sect.~\ref{sec:comp_sfr}). 

\smallskip
\noindent 
To summarize, for J0843 and RECX~11 the accretion rates computed
here are similar to previous accretion rates based on UV-excess. 
This similarity indicates both that the methodology is
stable and that these two objects had accretion rates that were
consistent on time baselines of a few weeks, the time between the
\citet{InglebyCalvet:2013aa} spectra and those obtained here.

\smallskip
\noindent
On the other hand, the accretion rates for RECX 9 and RECX 5 measured 
from the UV-excess are discrepant with those measured from H$\alpha$.
If real and not the result of different methodologies, then these changes are consistent with the accretion variability
of 0.37 dex inferred from H$\alpha$ monitoring,
\citep{CostiganScholz:2012aa}, which may be correlated with stellar
rotation as the accretion flow is seen from different sides at
different phases in the period. This variability may instead indicate 
quasi-periodic bursts \citep[e.g.,][]{CodyHillenbrand:2017aa}. Changes in the accretion
rate and line equivalent widths and profiles are not necessarily
correlated because the excess accretion continuum emission is thought to be produced by the
 accretion shock, while the H line emission is likely produced in the
 accretion funnel flows \citep[see modeling of line profiles by][]{AlencarBouvier:2012aa}.

%%%%%%%%%%%%%%%%%%%%%%%%%%%%%%%%%%%%%%%%%%%%%%%%%%%%%
\begin{figure}%[ht]		
\centering			
\includegraphics[width=1.\linewidth]{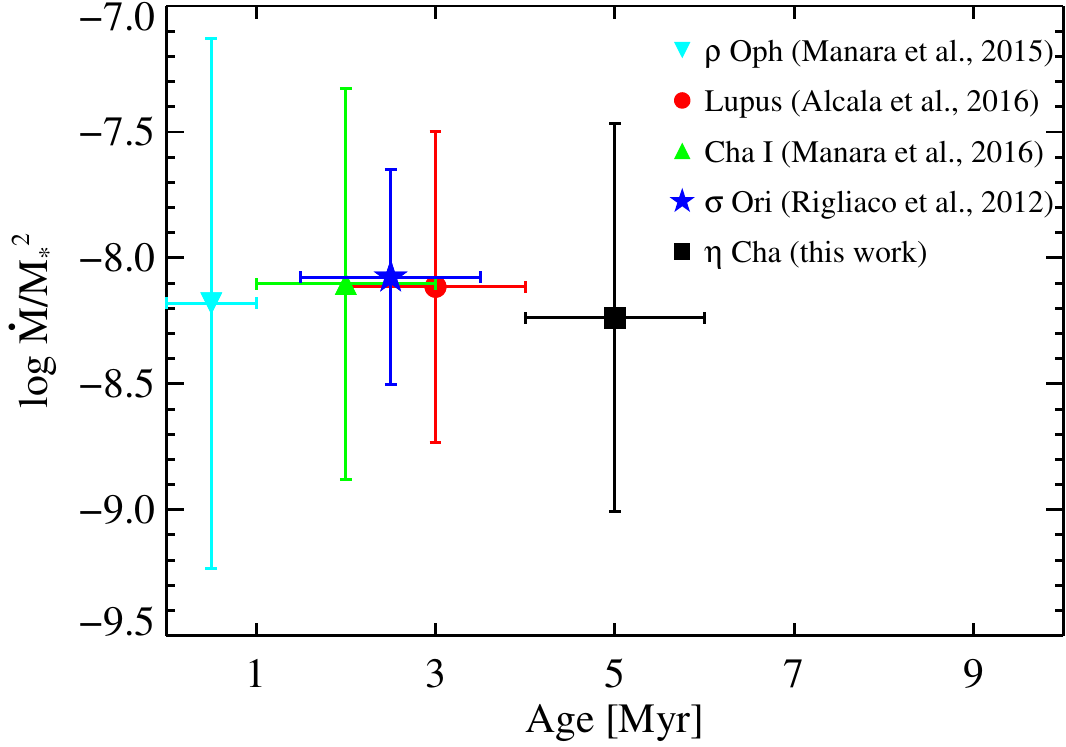}
\caption{Mass accretion rates normalized by stellar mass squared
at cluster ages of individual star forming regions. The mass accretion 
rate $\dot{M}$ of each star has been divided by $M_*^{2}$ and combined 
by the median for each cluster. $\dot{M}$ is given in units of ${\rm M_\sun/yr}$ and the stellar mass $M_{\rm *}$ is given in units of ${\rm M_\sun}$.
Stars in the $\eta$~Cha~association
are shown as a filled black square. The other clusters are shown as a
cyan inverted triangle for $\rho$ Ophiucus, as a red circle for Lupus,
as a green triangle for Chamaeleon I and as a blue star for $\sigma$
Orionis.  
The error bars indicate the standard deviation in logarithmic scale of 
the normalized mass accretion rates.}
\label{fig:plot_Mmacc_normed_age}
\end{figure}		

\begin{figure*}%[ht]		
\centering			
\includegraphics[width=.99\linewidth]{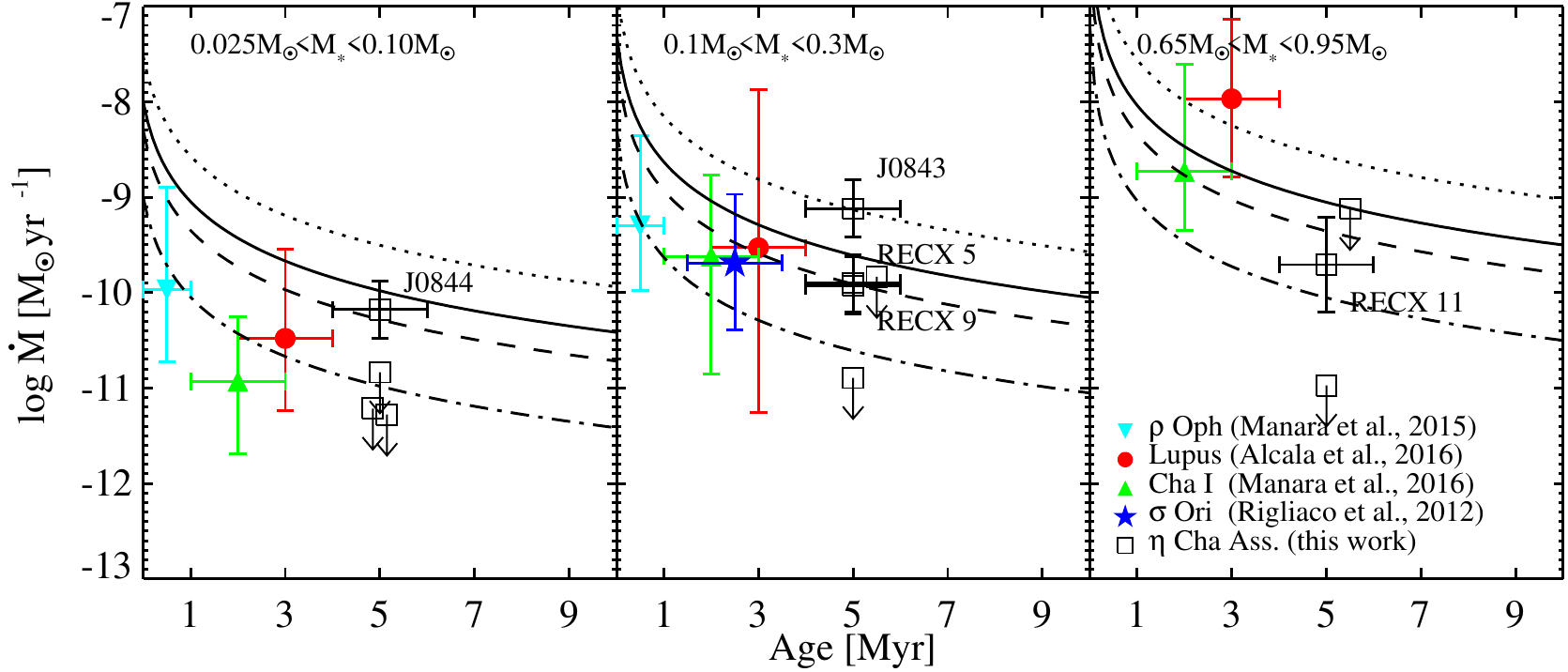}
\caption{Mass accretion rate versus cluster age for three different stellar
  mass ranges. The ranges are indicated in each panel. Stars in the $\eta$~Cha~association are shown as empty
  black squares. Upper limits are drawn as arrows for objects with no
  detected accretion, otherwise, colors and symbols as in
  Fig.~\ref{fig:plot_Mmacc_normed_age}. The dotted, solid, dashed and
  dashed-dotted lines indicate fiducial models for viscous accretion
  with disk masses of 30\%, 10\%, 5\% and 1\% of the mass of the
  central star, respectively. The stellar mass used in the model is
  $M_*=0.05\,{\rm M_\sun}$ in the \textit{left}, $M_*=0.15\,{\rm M_\sun}$ in
  the \textit{middle} and $M_*=0.8\,{\rm M_\sun}$ in the \textit{right panel} (see
  details and references in text).} 
\label{fig:plot_Mmacc_binned_age}
\end{figure*}		

\begin{figure}%[ht]	
\centering			
\includegraphics[width=1.\linewidth]{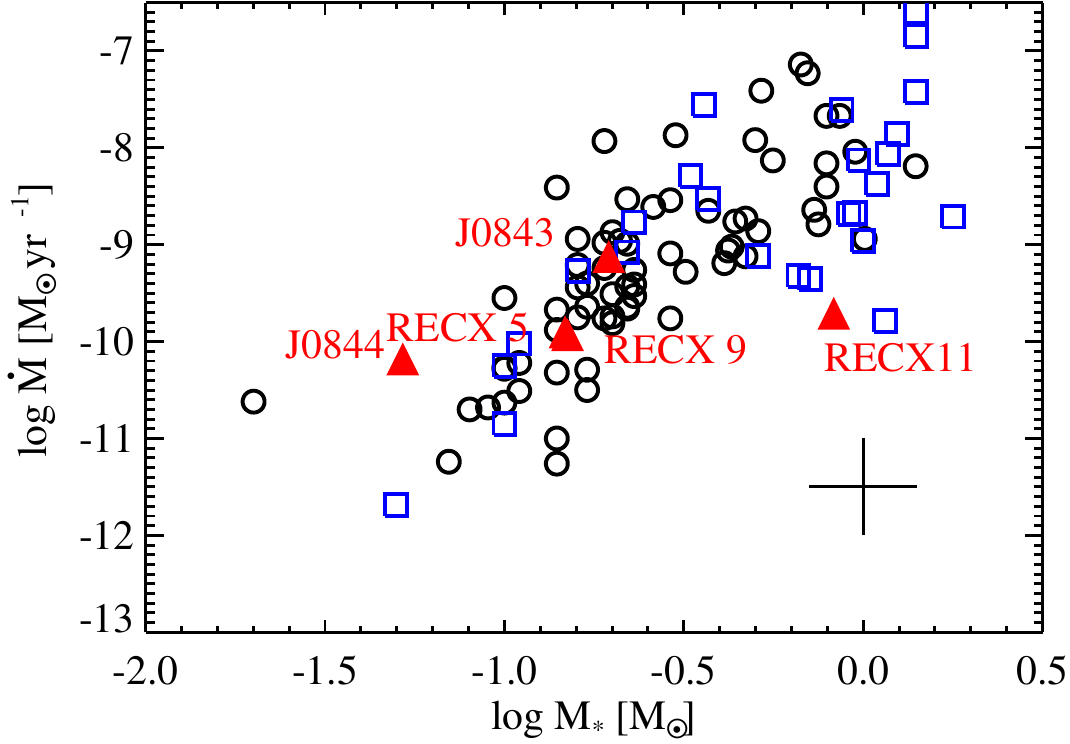}
\caption{Mass accretion rate vs. stellar mass for objects in the
  $\eta$~Cha cluster (red filled triangles; this work), the Lupus
  (black empty circles; \citealt{AlcalaManara:2017aa}) and Chamaeleon I
  star forming regions (blue empty squares;
  \citealt{ManaraFedele:2016aa}). Note that RECX~5 and RECX~9 overlap
  in this presentation, as they differ only slightly in stellar mass
  and mass accretion rate. In the \mbox{lower-right} corner, typical errors
  are indicated.} 
\label{fig:logm_logmacc}
\end{figure}		

\begin{figure}%[ht]
\centering			
\includegraphics[width=1.\linewidth]{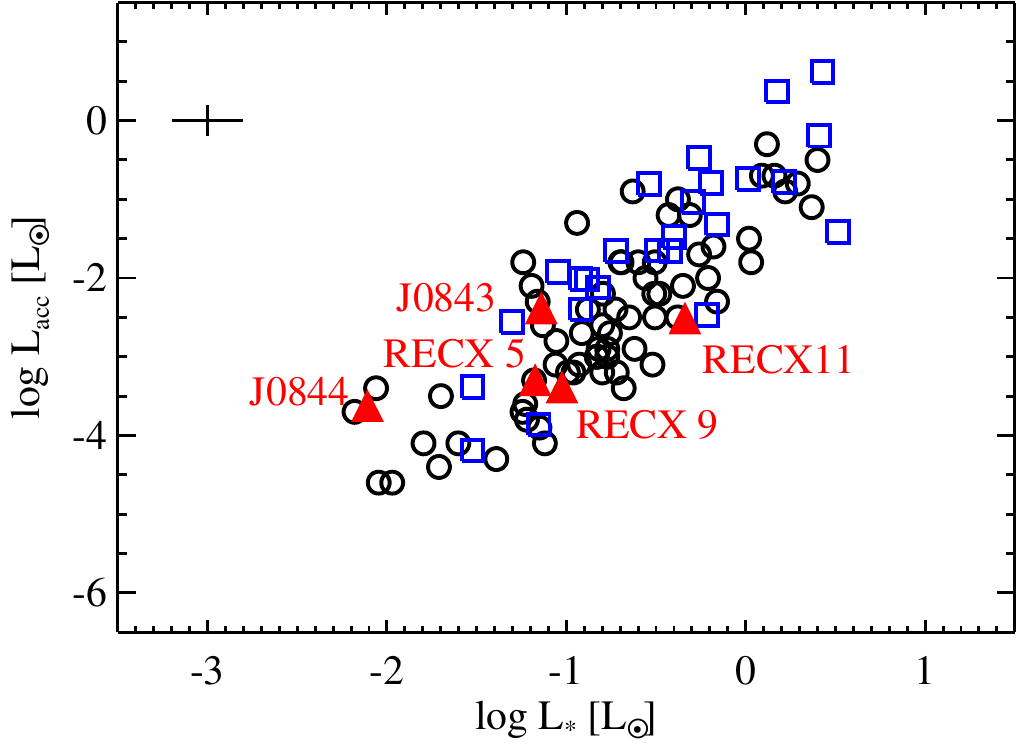}
\caption{Accretion luminosity vs. stellar mass for objects in the
  $\eta$~Cha cluster (red filled triangles; this work), the Lupus
  (black empty circles; \citealt{AlcalaManara:2017aa}) and Chamaeleon I
  star forming regions (blue empty squares;
  \citealt{ManaraFedele:2016aa}). In the \mbox{upper-left} corner, typical
  errors are indicated.} 
\label{fig:logl_loglacc}
\end{figure}		

\subsection{Comparisons to other low mass star forming regions} \label{sec:comp_sfr}
Figures \ref{fig:logm_logmacc} and \ref{fig:logl_loglacc}
compare the UV-excess measurements of $L_{\rm acc}$ and $\dot{M}$ of $\eta$ Cha
members with those measured onto stars in
the Lupus and Chamaeleon~I star forming regions
\citep{AlcalaNatta:2014aa,AlcalaManara:2017aa,ManaraFedele:2016aa}. The locus of
$\eta$ Cha objects is consistent with accretors in other regions. Therefore, the accretion 
properties appear to be similar. 

\smallskip
\noindent
To support this conclusion, we compare the mass accretion rates 
of the $\eta$~Cha association to those of other clusters with different ages. The
measurements have been compiled from the literature for the
$\rho$ Ophiucus \citep{ManaraTesti:2015aa}, Lupus
\citep{AlcalaManara:2017aa}, Chamaeleon I \citep{ManaraFedele:2016aa}
and $\sigma$ Orionis \citep{RigliacoNatta:2011aa} star forming
regions. These clusters have been selected because the mass accretion
rates were determined with a similar technique (except for $\rho$
Ophiucus, for which $\dot{M}$ was determined from
emission lines; \citealt{ManaraTesti:2015aa}) and the observations
were conducted with the same instrument (X-Shooter). We adopt cluster ages of 0.5~Myr for $\rho$ Ophiucus
\citep{GreeneMeyer:1995aa,LuhmanRieke:1999aa,MohantyJayawardhana:2005aa},
2~Myr for Chamaeleon I, 2.5~Myr for $\sigma$~Orionis \citep[both][and
references therein]{Fangvan-Boekel:2013aa} and 3~Myr for Lupus
\citep{AlcalaNatta:2014aa,AlcalaManara:2017aa}, and 5 Myr for the
$\eta$ Cha Association. The accretion rate
for each star is normalized by the relationship $\dot{M}
\propto M_*^{2.0}$ \citep[based on previous estimates with an exponent $\sim 2.0$ from, e.g.,][]{HartmannHerczeg:2016aa,ManaraTesti:2017aa}.

\smallskip
\noindent
The median mass accretion rates agree well for all clusters, despite differences in
age (Figure~\ref{fig:plot_Mmacc_normed_age}).  Disk dispersal is expected to be governed by viscous accretion at
least in some phases of its evolution, which implies a decrease of
mass accretion rate with time
\citep[e.g.,][]{HartmannCalvet:1998aa,AlexanderPascucci:2014aa}.  Among
stars with ongoing accretion, the decrease in accretion rate with time is not
detected.  

\smallskip
\noindent
However, drawing conclusions on the physical description of
disk accretion from these comparisons is challenging for several
reasons \citep[see
also,][]{ManaraFedele:2016aa,HartmannHerczeg:2016aa}: first, while
the average accretion rate may not change, the fraction of stars with
disks and with ongoing accretion decreases with time
\citep[e.g.][]{HaischLada:2001aa,HernandezHartmann:2008aa,Fedelevan-den-Ancker:2010aa}; second, stars in
a single cluster may have an age spread, in which case the stars with
disks may be younger than the average cluster age; finally, according to
models of viscous accretion, the change in accretion rate with stellar age will
flatten out, i.e., will decrease with stellar age
\citep[e.g.,][]{HartmannCalvet:1998aa}. The $\eta$~Chamaeleontis
cluster may therefore be biased towards higher initial disk masses, as
these live longer than their lower mass counterparts. Comparisons to
clusters at older ages to younger ones is therefore difficult
\citep{Sicilia-AguilarHenning:2010aa}. 
This effect may be enhanced by
the possible onset of photo-evaporation, which could quickly remove
disks once the mass accretion rate drops below about
$10^{-10}{\us}{\rm M_\sun/yr}$ for solar-mass stars
\citep{ClarkeGendrin:2001aa,GortiDullemond:2009aa}. 

\smallskip
\noindent
We therefore investigate how accretion rates of similar-mass stars compare among clusters
of different age, and how they relate to models of viscous evolution (Figure~\ref{fig:plot_Mmacc_binned_age}). 
As the $\eta$~Cha association hosts only five accreting objects, which span an order of magnitude in stellar
mass, the cluster samples are divided into three mass bins and each $\eta$~Cha object is drawn separately. 
The mass ranges have been chosen around the mass of J0844 ($0.025<M_\sun<0.1$;
left panel), of J0843, RECX~5 and RECX~9
($0.1<M_\sun<0.3$; middle panel), and of RECX~11 ($0.65<M_\sun<0.95$;
right panel). For the clusters, the mean in logarithmic scale of $\dot{M}$ is given. 
It is only shown, if a single mass bin contained at
least 3 objects. The error bars of $\dot{M}$ are the minimum
and maximum accretion rate in each mass bin. While upper limits
on $\dot{M}$ are drawn as arrows for non-accreting objects in the $\eta$~Cha
association, upper limits have not been included
for other clusters.  

\smallskip
\noindent
Fiducial models of accretion by viscous evolution
\citep{HartmannCalvet:1998aa} are for stellar masses of
$0.05$, $0.15$, and $0.8$ M$_\sun$, with initial disk masses of 30\%,
10\%, 5\% and 1\% times the mass of the central star. This simple
parametric model assumes $\alpha = 10^{-2}$, $R_1 = 10 {\rm~AU}$ and
$T_{\rm 100 AU} = 10{\us}{\rm K}$ (see \citealt{HartmannCalvet:1998aa}
for more details on the parameters). 

\smallskip
\noindent
In the lowest mass bin, the time evolution of accretion between $\rho$ Ophiucus and Lupus and
Chamaeleon I is traced well by a viscous accretion model with
$M_*=0.05\,M_\sun$ and a disk mass of $M_{\rm disk}=0.005\,M_\sun$
(dashed-dotted model in the left panel of
Fig.~\ref{fig:plot_Mmacc_binned_age}).  At the age of the $\eta$~Cha
cluster, mass accretion rates expected from this model would be below the detection limit of the method
presented here, as indicated by the upper limits of non-accreting
objects.  On the other hand, $\dot{M}$ of J0844
lies on the high end of the range of mass accretion rates found in the
Lupus and Chamaeleon I, which could be explained if J0844 had a high
initial disk mass.  J0843 also shows an enhanced accretion rate.

\smallskip
\noindent
RECX~5 and RECX~9 are accreting at rates that are lower
but within the range of accretion rates in Lupus, Chamaeleon I, and $\sigma$ Orionis. This difference is
in agreement with viscous accretion models. 
RECX~11 shows significantly lower $\dot{M}$ than seen in
other clusters for stars of similar mass. In direct comparison to
Chamaeleon I, this difference is in agreement with expectations from simple models of viscous
disk accretion (see 
Fig.~\ref{fig:plot_Mmacc_binned_age}). 

%%%%%%%%%%%%%%%%%%%%%%%%%%%%%%%%%%%%%%%%%%%%%%
% Conclusion
%%%%%%%%%%%%%%%%%%%%%%%%%%%%%%%%%%%%%%%%%%%%%%
\section{Conclusions}
We present a revised analysis of the mass accretion and stellar
properties of 15 low-mass stars in the nearby $\eta$~Cha
association. Thanks to the simultaneous, \mbox{broad-wavelength-range}
coverage and \mbox{flux-calibration} accuracy of VLT/X-Shooter we determined
the spectral type, extinction and accretion luminosity in a
\mbox{self-consistent} way. Out of the 15 low-mass stars studied here, we
detected ongoing mass accretion in 5 systems. Once compared with
literature values, we find that the mass accretion rates are
consistent within errors with previous studies for which UV excess
measurements exist, and deviating for most objects, when comparing to results from H$\alpha$
modeling due to methodological differences. We also
report a mass accretion rate for J0844, for which no direct
measurement of mass accretion has been reported in the literature. The
derived mass accretion rates in the $\eta$~Cha cluster are similar to
the values measured in younger star forming regions. 

\begin{acknowledgements}
We kindly thank the anonymous referee for the careful read, the detailed suggestions and thorough comments, 
which helped to significantly improve the readability and to strengthen the scientific message of the paper.
We thank C. F. Manara for kindly providing us the data of
non-accreting PMS and for discussion. We also thank Paula Teixeira for
help with the observations and together with Kevin Covey and
Adam Kraus for help in preparing the proposal.
MR is a fellow of the International Max Planck Research School
for Astronomy and Cosmic Physics (IMPRS) at the University of
Heidelberg. DF acknowledges support from the Italian Ministry of Education, Universities and Research project SIR (RBSI14ZRHR). 
GJH is supported by general grant 11473005 awarded by the National Science Foundation of China. This research has made use of the VizieR
catalogue access tool, CDS, Strasbourg, France.
\end{acknowledgements}

%%%%%%%%%%%%%%%%%%%%%%%%%%%%%%%%%%%%%%%%%%%%%%
% Bibliography
%%%%%%%%%%%%%%%%%%%%%%%%%%%%%%%%%%%%%%%%%%%%%%
\bibliographystyle{aa}
\bibliography{Library}

%%%%%%%%%%%%%%%%%%%%%%%%%%%%%%%%%%%%%%%%%%%%
% Appendix
%%%%%%%%%%%%%%%%%%%%%%%%%%%%%%%%%%%%%%%%%%%%
\clearpage
\begin{appendix}

%%%%%%%%%%%%%%%%%%%%%%%%%%%%%%%%%%%%%%%%%%%%
\section{Spectra and best-fit models of accretors in the $\eta$~Cha cluster}
% J0843
\begin{figure*}	
\centering			
\includegraphics[width=.85\linewidth]{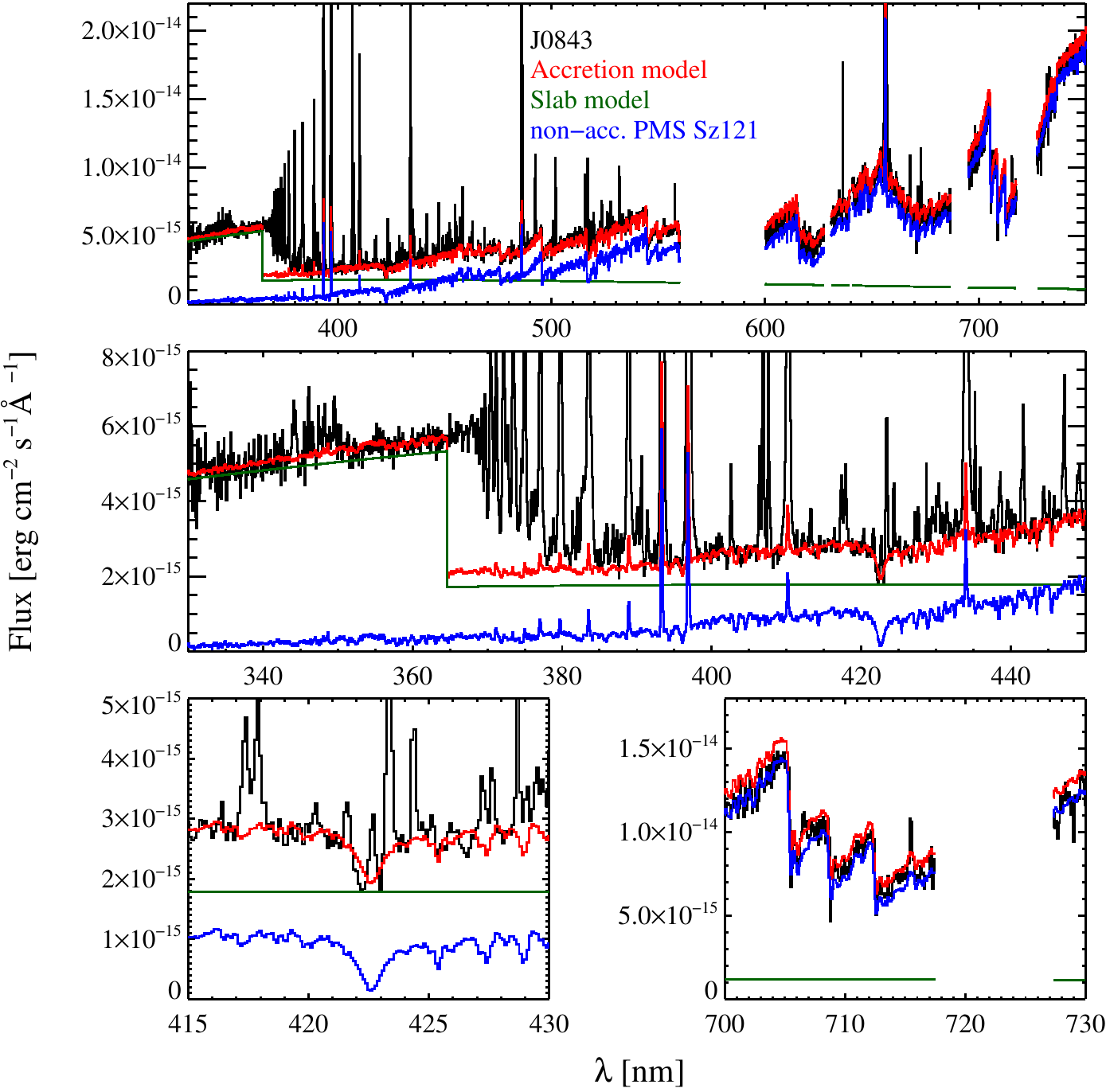}
\caption{Accretion model (red) of PMS J0843 (black), using as basis the non-accreting PMS Sz121 \citep[blue;][]{ManaraTesti:2013aa} and a plane parallel Hydrogen slab \citep[green;][]{ValentiBasri:1993aa}. The spectra have been re-binned to 0.1 nm resolution.}
\label{fig:plot_w_models_s015}
\end{figure*}		

% J0844
\begin{figure*}	
\centering			 
\includegraphics[width=.85\linewidth]{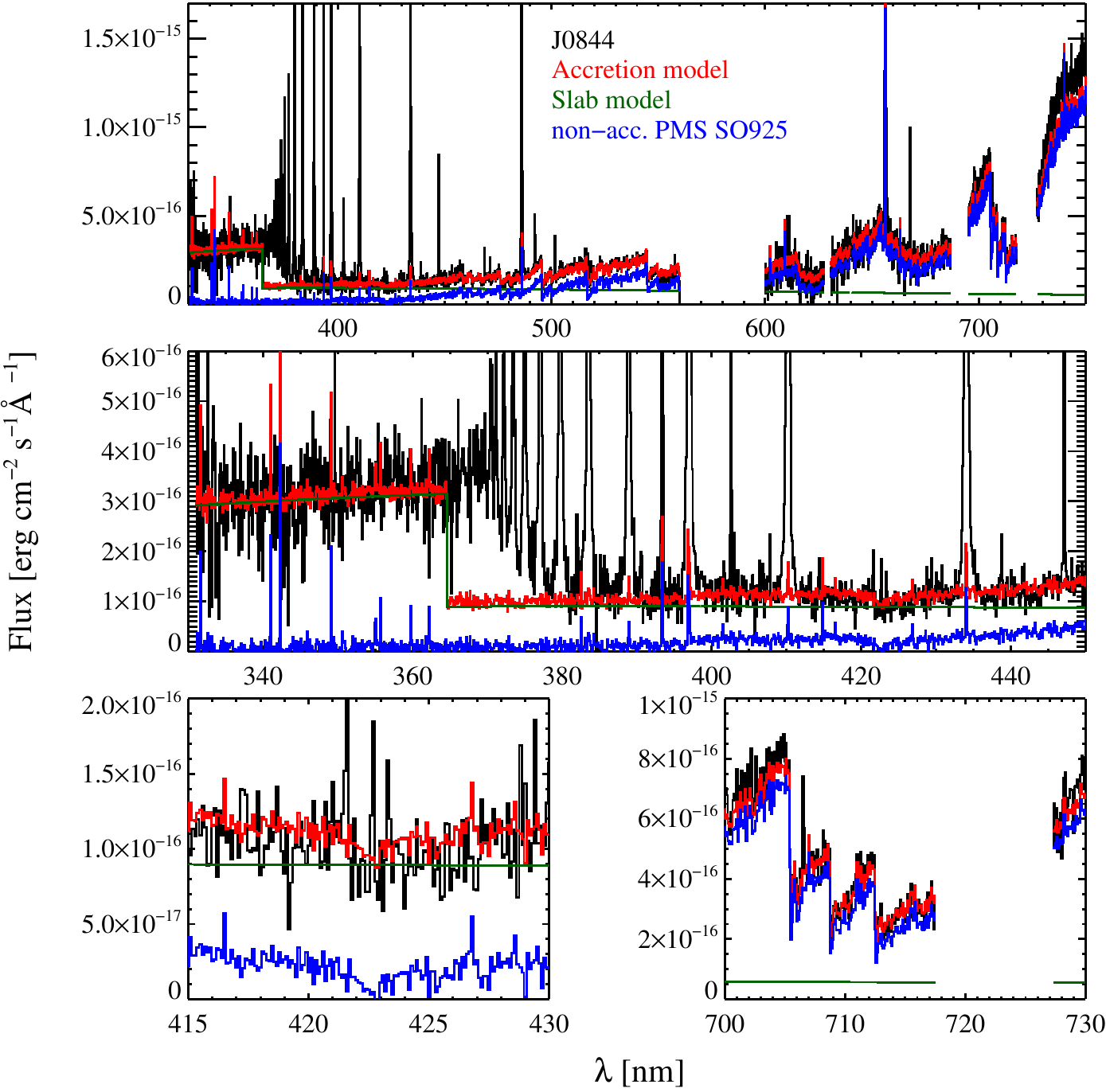}
\caption{Accretion model of PMS J0844, using as basis the non-accreting PMS SO925 \citep{ManaraTesti:2013aa}. Colors as in Fig.~\ref{fig:plot_w_models_s015}.}
\label{fig:plot_w_models_s016}
\end{figure*}		

% RECX~5
\begin{figure*}	
\centering			
\includegraphics[width=.85\linewidth]{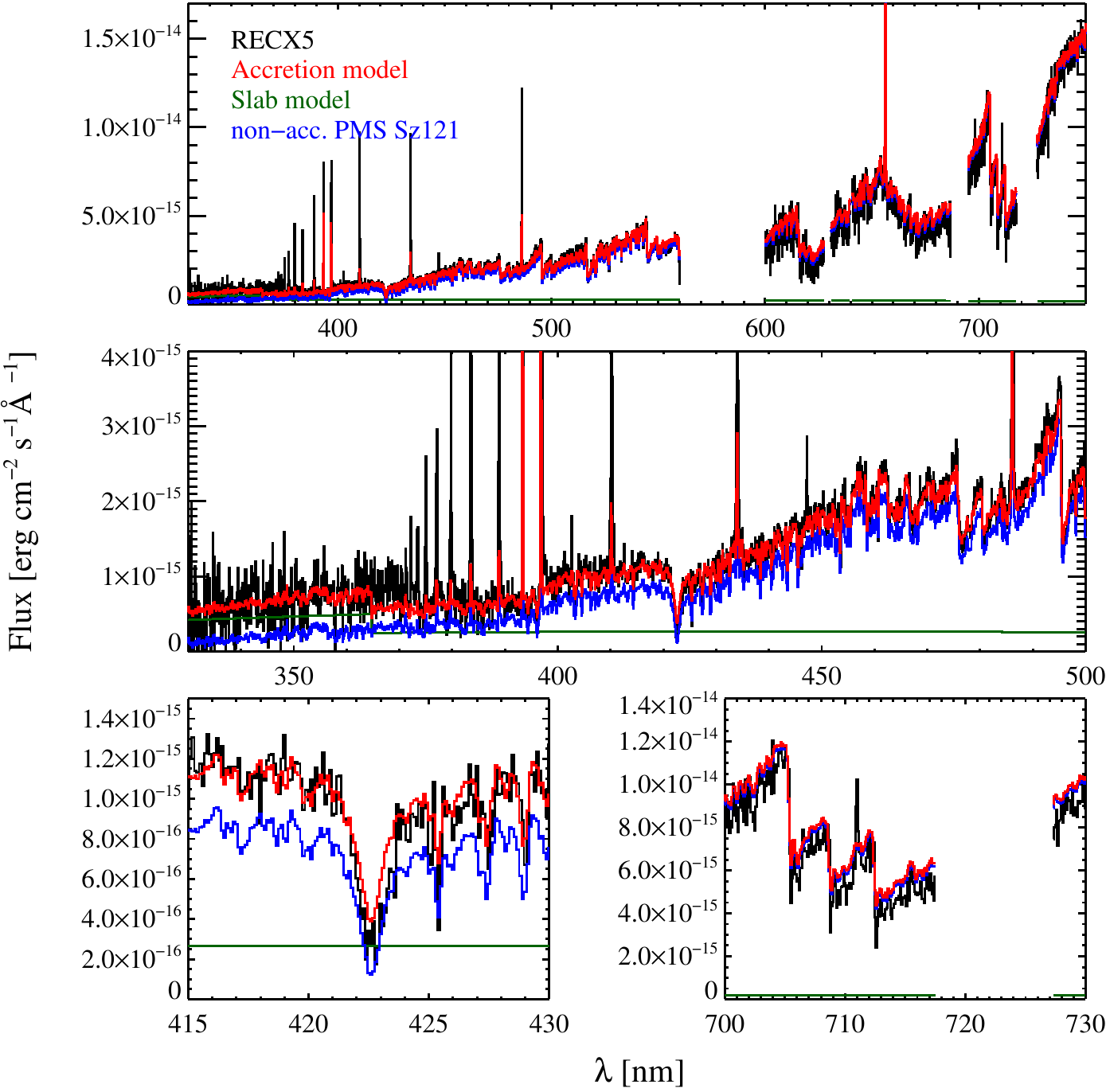}
\caption{Accretion model to PMS RECX5, using as basis the non-accreting PMS Sz121 \citep{ManaraTesti:2013aa}. Colors as in Fig.~\ref{fig:plot_w_models_s015}.}
\label{fig:plot_w_models_s041}
\end{figure*}		

% RECX~9
\begin{figure*}	
\centering			
\includegraphics[width=.85\linewidth]{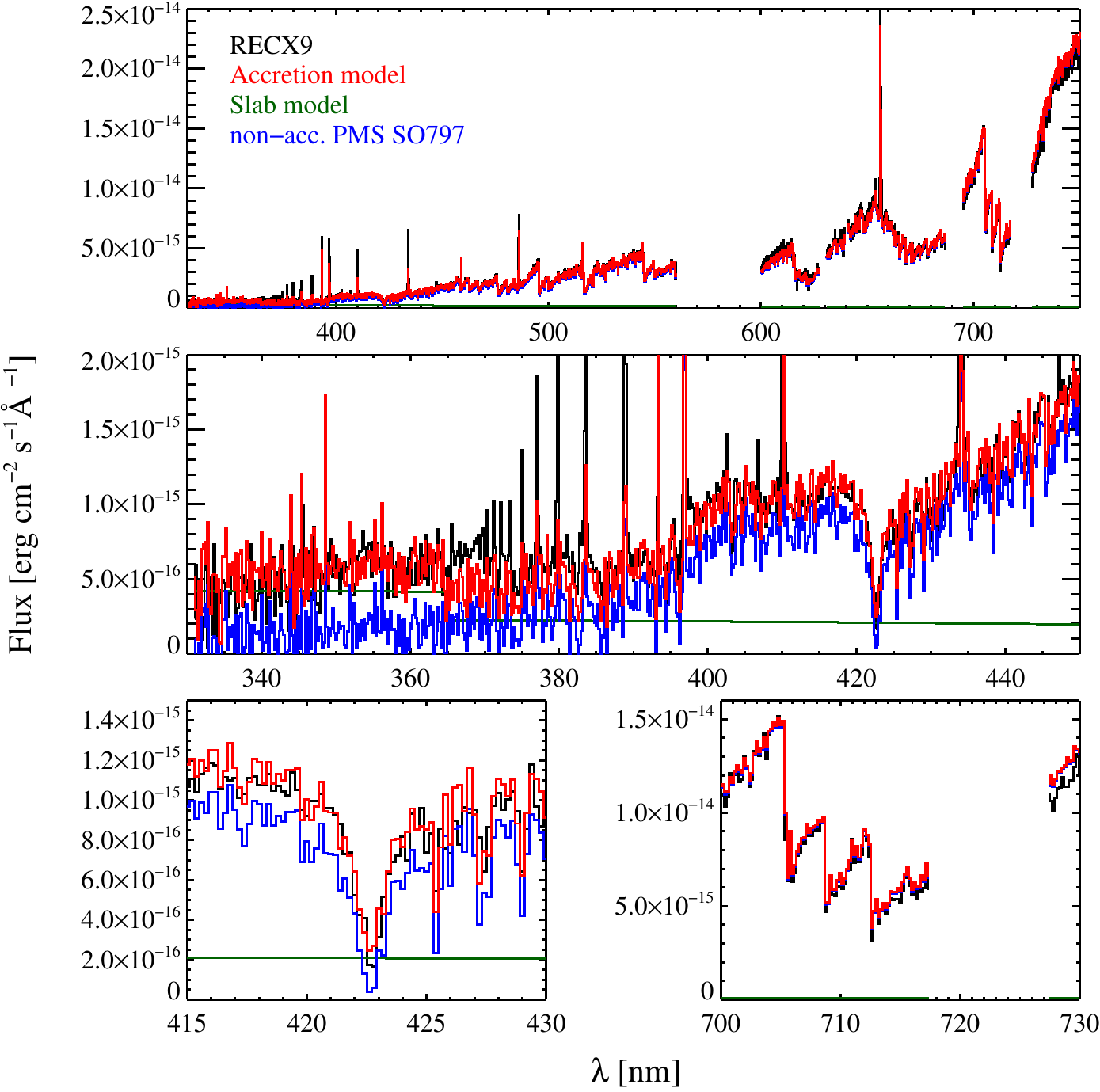}
\caption{Accretion model to PMS RECX~9, using as basis the non-accreting PMS SO797 \citep{ManaraTesti:2013aa}. Colors as in Fig.~\ref{fig:plot_w_models_s015}. The spectra have been re-binned to 0.2 nm resolution.
}
\label{fig:plot_w_models_s042}
\end{figure*}		

% RECX~11
\begin{figure*}	%[ht]	
\centering			
\includegraphics[width=.85\linewidth]{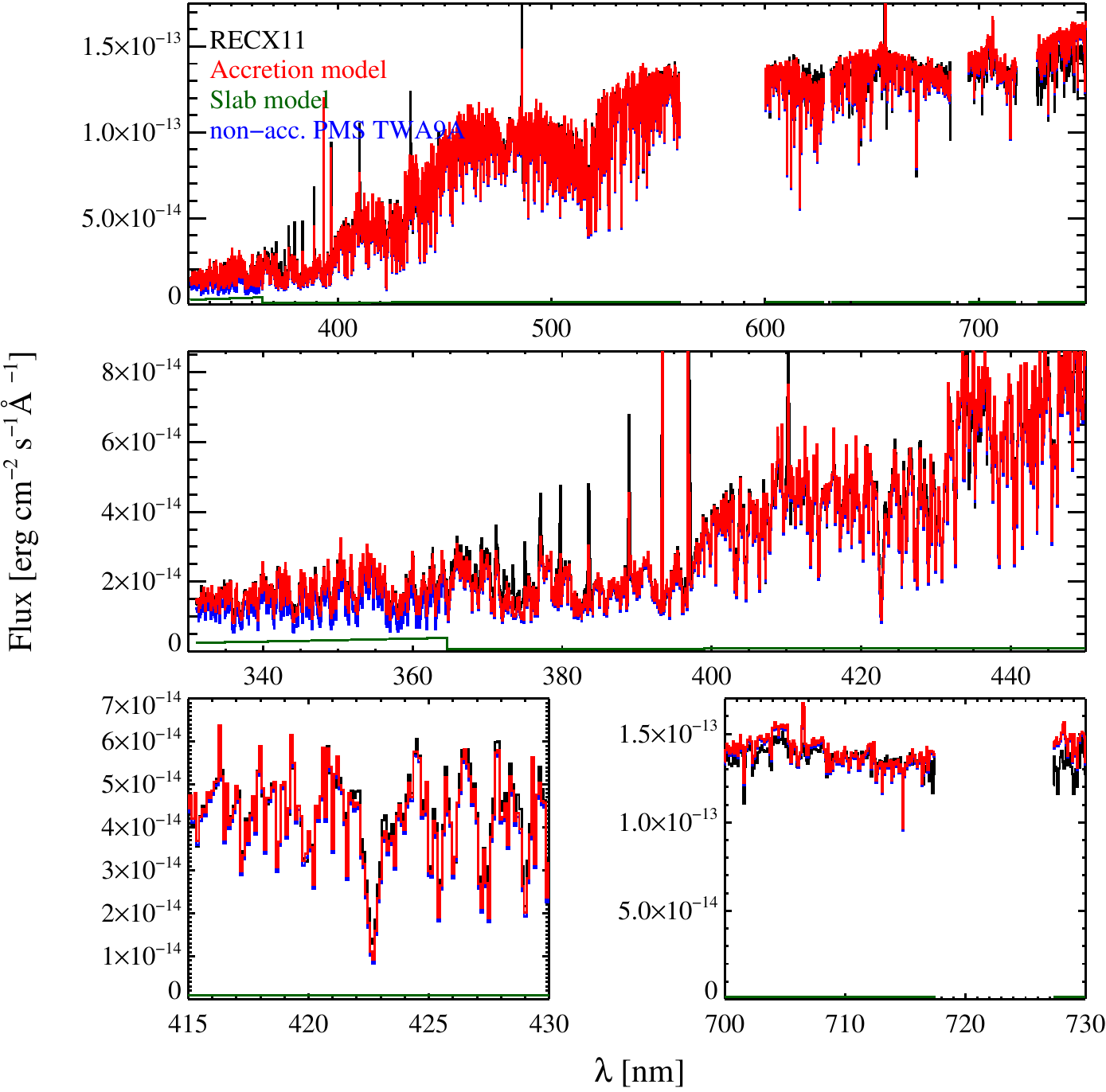}
\caption{Accretion model to PMS RECX~11, using as basis the non-accreting PMS TWA9A \citep{ManaraTesti:2013aa}. Colors as in Fig.~\ref{fig:plot_w_models_s015}.}
\label{fig:plot_w_models_s011}
\end{figure*}		

In Figures \ref{fig:plot_w_models_s015}-\ref{fig:plot_w_models_s011}, the accreting PMSs in the $\eta$~Cha cluster are shown together with the models, which were used to determine the accretion flux.

\clearpage
%%%%%%%%%%%%%%%%%%%%%%%%%%
%Plots
\section{Emission lines} \label{app:sec_appendixb}
We report flux and equivalent width of the H$\alpha$ and H$\beta$ transitions, the \ion{He}{i}${~\lambda 402.6}$, ${{\rm HeI}~\lambda 447.1}$, \ion{He}{i}${~\lambda 501.5}$, \ion{He}{i}${~\lambda 587.5}$, \ion{He}{i}${~\lambda 667.8}$, \ion{He}{i}${~\lambda 706.5}$ transitions and the ${\rm HeIFeI}\lambda 492.2$ emission feature \citep{AlcalaNatta:2014aa}. We include the detections of the [\ion{O}{i}]${~\lambda 557.7}$, [\ion{O}{i}]${~\lambda 630.0} $ and [\ion{O}{i}]${~\lambda 636.3}$ forbidden transitions. 

% Oxygen transitions
\begin{table*}
\centering	
\caption{
Fluxes and equivalent widths of forbidden Oxygen transitions
}
\begin{tabular}{l|r|r|r|r|r|r} 
 \hline 
 \hline 
 Object & $f_{ {\rm [OI]}~\lambda 557.7} $ &$EW_{{\rm [OI]}~\lambda 557.7} $ &$f_{ {\rm [OI]}~\lambda 630.0} $ &$EW_{{\rm [OI]}~\lambda 630.0} $ &$f_{ {\rm [OI]}~\lambda 636.3} $ &$EW_{{\rm [OI]}~\lambda 636.3} $ \\
  & [${\rm erg} {\rm s}^{-1}{\rm cm}^{-2} $] &[\AA] &[${\rm erg} {\rm s}^{-1}{\rm cm}^{-2} $] &[\AA] &[${\rm erg} {\rm s}^{-1}{\rm cm}^{-2} $] &[\AA] \\
 \hline 
J0836&<9.56e-18&-&<2.92e-17&-&<5.84e-17&-\\
RECX1&<1.24e-14&-&<9.35e-15&-&<6.00e-15&-\\
J0838&<3.63e-17&-&<1.26e-16&-&<1.72e-16&-\\
J0841&<1.30e-17&-&6.65($\pm$0.93)e-16&-1.18$\pm$0.18&<6.42e-17&-\\
RECX3&<1.62e-16&-&<3.61e-16&-&<5.48e-16&-\\
RECX4&<9.36e-16&-&<1.14e-15&-&<1.08e-15&-\\
RECX5&9.34($\pm$3.45)e-16&-0.27$\pm$0.10&<3.77e-16&-&<4.85e-16&-\\
RECX6&<3.03e-16&-&<3.26e-16&-&<4.46e-16&-\\
RECX7&<5.86e-15&-&<5.63e-15&-&<9.11e-15&-\\
J0843&6.94($\pm$0.47)e-15&-1.27$\pm$0.08&7.18($\pm$0.32)e-14&-12.91$\pm$0.48&2.40($\pm$0.20)e-14&-3.62$\pm$0.35\\
J0844&1.35($\pm$0.51)e-16&-0.73$\pm$0.30&<4.04e-17&-&<6.13e-17&-\\
RECX9&1.64($\pm$0.29)e-15&-0.47$\pm$0.09&<2.98e-16&-&<3.95e-16&-\\
RECX10&<1.66e-15&-&<1.86e-15&-&<1.59e-15&-\\
RECX11&<5.16e-15&-&<4.62e-15&-&<2.37e-15&-\\
\hline 
\end{tabular} 
\label{tbl:emission_1}
\end{table*}

% Helium transitions (1)
\begin{table*}
\centering	
\caption{
Fluxes and equivalent widths of Helium transitions (1)
}
\begin{tabular}{l|r|r|r|r|r|r} 
 \hline 
 \hline 
 Object & $f_{ {\rm HeI}~\lambda 402.6} $ &$EW_{{\rm HeI}~\lambda 402.6} $ &$f_{ {\rm HeI}~\lambda 447.1} $ &$EW_{{\rm HeI}~\lambda 447.1} $ &$f_{ {\rm HeI}~\lambda 501.5} $ &$EW_{{\rm HeI}~\lambda 501.5} $ \\
  & [${\rm erg} {\rm s}^{-1}{\rm cm}^{-2} $] &[\AA] &[${\rm erg} {\rm s}^{-1}{\rm cm}^{-2} $] &[\AA] &[${\rm erg} {\rm s}^{-1}{\rm cm}^{-2} $] &[\AA] \\
 \hline 
J0836&<1.54e-17&-&1.08($\pm$0.23)e-16&-0.74$\pm$0.17&<1.07e-17&-\\
RECX1&<1.38e-14&-&<1.27e-14&-&<1.40e-14&-\\
J0838&<4.74e-17&-&<3.29e-17&-&<2.40e-17&-\\
J0841&9.42($\pm$4.74)e-17&-0.61$\pm$0.32&<1.32e-17&-&<1.65e-17&-\\
RECX3&<1.87e-16&-&<1.46e-16&-&<2.25e-16&-\\
RECX4&<5.80e-16&-&<7.96e-16&-&<1.03e-15&-\\
RECX5&1.73($\pm$0.34)e-15&-2.10$\pm$0.49&2.22($\pm$0.32)e-15&-1.33$\pm$0.22&<9.46e-17&-\\
RECX6&<2.30e-16&-&<1.71e-16&-&<2.91e-16&-\\
RECX7&<7.74e-15&-&<8.81e-15&-&<6.56e-15&-\\
J0843&6.01($\pm$0.49)e-15&-2.34$\pm$0.19&1.13($\pm$0.06)e-14&-3.09$\pm$0.14&2.34($\pm$0.11)e-14&-5.69$\pm$0.22\\
J0844&6.66($\pm$0.64)e-16&-5.08$\pm$0.63&1.28($\pm$0.07)e-15&-8.08$\pm$0.71&3.30($\pm$0.31)e-16&-2.27$\pm$0.24\\
RECX9&9.03($\pm$1.99)e-16&-0.88$\pm$0.20&1.33($\pm$0.17)e-15&-0.81$\pm$0.10&<9.75e-17&-\\
RECX10&<8.34e-16&-&<1.26e-15&-&<1.62e-15&-\\
RECX11&<5.13e-15&-&<5.13e-15&-&<6.05e-15&-\\
RECX12&2.68($\pm$0.63)e-15&-0.52$\pm$0.13&4.69($\pm$1.23)e-15&-0.42$\pm$0.11&<5.55e-16&-\\
 \hline 
\end{tabular} 
\label{tbl:emission_2}
\end{table*}

% Helium transitions (2)
\begin{table*}
\centering	
\caption{
Fluxes and equivalent widths of Helium transitions (2)
}
\begin{tabular}{l|r|r|r|r|r|r} 
 \hline 
 \hline 
 Object & $f_{ {\rm HeI}~\lambda 587.5} $ &$EW_{{\rm HeI}~\lambda 587.5} $ &$f_{ {\rm HeI}~\lambda 667.8} $ &$EW_{{\rm HeI}~\lambda 667.8} $ &$f_{ {\rm HeI}~\lambda 706.5} $ &$EW_{{\rm HeI}~\lambda 706.5} $ \\
  & [${\rm erg} {\rm s}^{-1}{\rm cm}^{-2} $] &[\AA] &[${\rm erg} {\rm s}^{-1}{\rm cm}^{-2} $] &[\AA] &[${\rm erg} {\rm s}^{-1}{\rm cm}^{-2} $] &[\AA] \\
 \hline 
J0836&<7.02e-17&-&<6.24e-17&-&<9.36e-17&-\\
RECX1&<1.08e-14&-&<3.21e-15&-&<4.20e-15&-\\
J0838&<2.15e-16&-&<1.64e-16&-&<1.78e-16&-\\
J0841&<9.16e-17&-&<8.71e-17&-&<9.63e-17&-\\
RECX3&<5.20e-16&-&<4.37e-16&-&<9.34e-16&-\\
RECX4&<1.52e-15&-&<8.19e-16&-&<1.87e-15&-\\
RECX5&4.64($\pm$1.29)e-15&-1.86$\pm$0.51&3.97($\pm$1.33)e-15&-0.95$\pm$0.38&<4.67e-16&-\\
RECX6&<5.48e-16&-&<3.28e-16&-&<6.29e-16&-\\
RECX7&<6.53e-15&-&<1.71e-15&-&<3.19e-15&-\\
J0843&2.25($\pm$0.19)e-14&-4.70$\pm$0.35&6.37($\pm$1.24)e-15&-0.85$\pm$0.19&<4.91e-16&-\\
J0844&1.45($\pm$0.22)e-15&-9.95$\pm$1.44&1.00($\pm$0.13)e-15&-4.66$\pm$1.13&6.17($\pm$1.04)e-16&-1.56$\pm$0.28\\
RECX9&4.76($\pm$1.03)e-15&-1.66$\pm$0.35&<2.51e-16&-&<4.02e-16&-\\
RECX10&<1.97e-15&-&<9.02e-16&-&<1.84e-15&-\\
RECX11&<5.20e-15&-&<1.53e-15&-&<2.24e-15&-\\
RECX12&1.43($\pm$0.19)e-14&-0.70$\pm$0.09&<7.17e-16&-&<1.52e-15&-\\
\hline 
\end{tabular} 
\label{tbl:emission_3}
\end{table*}

% Helium transitions (3)
\begin{table*}
\centering	
\caption{
Fluxes and equivalent widths of Helium (3), H$\alpha$ and H$\beta$ transitions
}
\begin{tabular}{l|r|r|r|r|r|r} 
 \hline 
 \hline 
 Object & $f_{ {\rm HeIFeI}~\lambda 492.2} $ &$EW_{{\rm HeIFeI}~\lambda 492.2} $ &$f_{ {\rm H\alpha}~\lambda 656.3} $ &$EW_{{\rm H\alpha}~\lambda 656.3} $ &$f_{ {\rm H\beta}~\lambda 486.1} $ &$EW_{{\rm H\beta}~\lambda 486.1} $ \\
  & [${\rm erg} {\rm s}^{-1}{\rm cm}^{-2} $] &[\AA] &[${\rm erg} {\rm s}^{-1}{\rm cm}^{-2} $] &[\AA] &[${\rm erg} {\rm s}^{-1}{\rm cm}^{-2} $] &[\AA] \\
 \hline
 J0836&<1.36e-17&-&1.25($\pm$0.06)e-14&-13.38$\pm$0.72&2.10($\pm$0.09)e-15&-11.13$\pm$0.37\\
RECX1&<1.17e-14&-&3.19($\pm$0.27)e-13&-1.05$\pm$0.09&<1.22e-14&-\\
J0838&<2.66e-17&-&2.23($\pm$0.12)e-14&-11.98$\pm$0.77&4.19($\pm$0.20)e-15&-10.42$\pm$0.40\\
J0841&<1.85e-17&-&1.10($\pm$0.06)e-14&-9.45$\pm$0.62&2.32($\pm$0.11)e-15&-8.26$\pm$0.29\\
RECX3&<2.13e-16&-&3.94($\pm$0.38)e-14&-2.73$\pm$0.28&7.97($\pm$0.61)e-15&-1.91$\pm$0.14\\
RECX4&<9.90e-16&-&1.85($\pm$0.11)e-13&-4.24$\pm$0.26&4.50($\pm$0.30)e-14&-2.68$\pm$0.16\\
RECX5&<8.64e-17&-&8.68($\pm$0.42)e-14&-13.50$\pm$0.68&2.92($\pm$0.12)e-14&-14.70$\pm$0.35\\
RECX6&<2.80e-16&-&8.93($\pm$0.47)e-14&-5.32$\pm$0.24&2.28($\pm$0.11)e-14&-4.03$\pm$0.14\\
RECX7&<7.28e-15&-&1.65($\pm$0.21)e-13&-0.79$\pm$0.10&<7.20e-15&-\\
J0843&2.40($\pm$0.11)e-14&-4.99$\pm$0.11&9.90($\pm$0.40)e-13&-94.45$\pm$2.63&1.75($\pm$0.07)e-13&-42.61$\pm$0.68\\
J0844&6.64($\pm$0.39)e-16&-3.26$\pm$0.16&4.40($\pm$0.18)e-14&-102.20$\pm$5.45&1.40($\pm$0.06)e-14&-82.07$\pm$1.98\\
RECX9&<9.69e-17&-&9.60($\pm$0.43)e-14&-12.60$\pm$0.47&2.10($\pm$0.09)e-14&-10.73$\pm$0.30\\
RECX10&<1.31e-15&-&9.14($\pm$1.05)e-14&-1.68$\pm$0.20&2.76($\pm$0.38)e-14&-1.15$\pm$0.16\\
RECX11&<5.27e-15&-&1.16($\pm$0.05)e-12&-8.32$\pm$0.29&2.56($\pm$0.21)e-13&-2.73$\pm$0.22\\
RECX12&<6.36e-16&-&3.05($\pm$0.13)e-13&-8.41$\pm$0.22&1.06($\pm$0.04)e-13&-8.15$\pm$0.16\\
\hline 
\end{tabular} 
\label{tbl:emission_4}
\end{table*}
\end{appendix}

%%%%%%%%%%%%%%%%%%%%%%%%%%%%%%%%%%%%%%%%%%%%%%%%%%%%%%%%%
%%%%%%%%%%% END BODY %%%%%%%%%%%%%%%%%%%%%%%%%%%%%%%%%%%%
%%%%%%%%%%%%%%%%%%%%%%%%%%%%%%%%%%%%%%%%%%%%%%%%%%%%%%%%%
\end{document}